\documentclass[twocolumn, a4paper,accepted=2020-09-15]{quantumarticle}
\pdfoutput=1

\usepackage{graphicx}
\usepackage{dcolumn}
\usepackage{bm}
\usepackage{epsfig}
\usepackage{amsmath}
\usepackage{amssymb}
\usepackage{color}
\usepackage{bbm}
\usepackage{braket}

\usepackage[colorlinks=true,citecolor=blue,linkcolor=blue,urlcolor=blue]{hyperref}

\begin{document}

\title{Thermodynamics of ultrastrongly coupled light-matter systems}

\author{Philipp Pilar$^1$, Daniele De Bernardis$^1$, and Peter Rabl$^1$}
\affiliation{$^1$Vienna Center for Quantum Science and Technology, Atominstitut, TU Wien, 1040 Vienna, Austria}

\begin{abstract} 
We study the thermodynamic properties of a system of two-level dipoles that are coupled ultrastrongly to a single cavity mode.  By using exact numerical and approximate analytical methods, we evaluate the free energy of this system at arbitrary interaction strengths and discuss strong-coupling modifications of derivative quantities such as the specific heat or the electric susceptibility. From this analysis we identify the lowest-order cavity-induced corrections to those quantities in the collective ultrastrong coupling regime and show that for even stronger interactions the presence of a single cavity mode can strongly modify extensive thermodynamic quantities of a large ensemble of dipoles. In this non-perturbative coupling regime we also observe a significant shift of the ferroelectric phase transition temperature and a characteristic broadening and collapse of the black-body spectrum of the cavity mode. Apart from a purely fundamental interest, these general insights will be important for identifying potential applications of ultrastrong-coupling effects, for example, in the field of quantum chemistry or for realizing quantum thermal machines.
\end{abstract}
 
\maketitle

%
%


\maketitle
\section{Introduction}
\label{sec:introduction}

Undoubtedly, the interplay between statistical physics and the theory of electromagnetic (EM) radiation played a very important role in the history of modern physics. Discrepancies between the predicted and the measured spectrum of black-body radiation led to the birth of quantum mechanics. Based on purely thermodynamic arguments, Einstein introduced his A-coefficient and postulated the effect of spontaneous emission, long before it was understood microscopically.  Investigations of photon-photon correlations from thermal and coherent sources of light stood at the beginning of the field of quantum optics, and so on. In most of these and related examples the EM field can be treated as an independent subsystem, which thermalizes via weak interactions with the surrounding matter. This assumption breaks down in the so-called ultrastrong coupling (USC) regime~\cite{Ciuti2005,Forndiaz2019,Kockum2019}, where the interaction energy can be comparable to the bare energy of the photons. Such conditions can be reached in solid-state~\cite{Todorov2010,Scalari2012,Dietze2013,Gubbin2014,Zhang2016,Bayer2017,Askenazi2017} and molecular cavity QED experiments~\cite{Schwartz2011,George2016,Flick2017,Ribeiro2018,Peters2019}, where modifications of chemical reactions~\cite{Hutchison2012,Thomas2016} or phase transitions~\cite{Wang2014} have  been observed and interpreted as vacuum-induced changes of thermodynamic potentials~\cite{CanaguierDurand2013}. Together with the ability to realize even stronger couplings between artificial superconducting atoms and microwave photons~\cite{Devoret2007,Niemczyk2010,Forndiaz2010,Forndiaz2017,Yoshihara2017}, these observations have led to a growing interest~\cite{Forndiaz2019,Kockum2019} in the ground and thermal states of light-matter systems under conditions where the coupling between the individual parts can no longer be neglected.

Since an exact theoretical treatment of light-matter systems in the  USC regime is in general not possible, one usually resorts to simplified descriptions, for example, based on the Dicke~\cite{Dicke1954,Brandes2005} or the Hopfield~\cite{Hopfield} model. However, such reduced models often do not represent the complete energy of the system~\cite{Rzazewski1975,ViehmannPRL2011,Todorov2012,Bamba2014,Jaako2016,DeBernardis2018,Rokaj2018,Andolina2019} or contain gauge artefacts~\cite{DeBernardis2018,DeBernardis2018b,Stokes2019,Stefano2019,Roth2019} that prevent their applicability in the USC regime. More generally, while in weakly coupled cavity QED systems the role of static dipole-dipole interactions can often be neglected or modelled independently of the dynamical EM mode, this is no longer the case in the USC regime~\cite{DeBernardis2018,Kudenko1975,Keeling2007,Vukic2012,Griesser2016,Stokes2020}. An inconsistent treatment of static and dynamical fields can thus very easily lead to wrong predictions or a misinterpretation of results. A prominent example in this respect is the superradiant phase transition of the Dicke model~\cite{Hepp1973,Wang1973,Carmichael1973}, which is often described as cavity-induced, but which can be understood as a regular ferroelectric instability in a system of strongly attractive dipoles~\cite{DeBernardis2018,Keeling2007}. In the past, these and other subtle issues have led to many controversies in this field and prevented a detailed understanding of the ground- and thermal states of USC light-matter systems so far.

\begin{figure}[b]
\centering
	\includegraphics[width=\columnwidth]{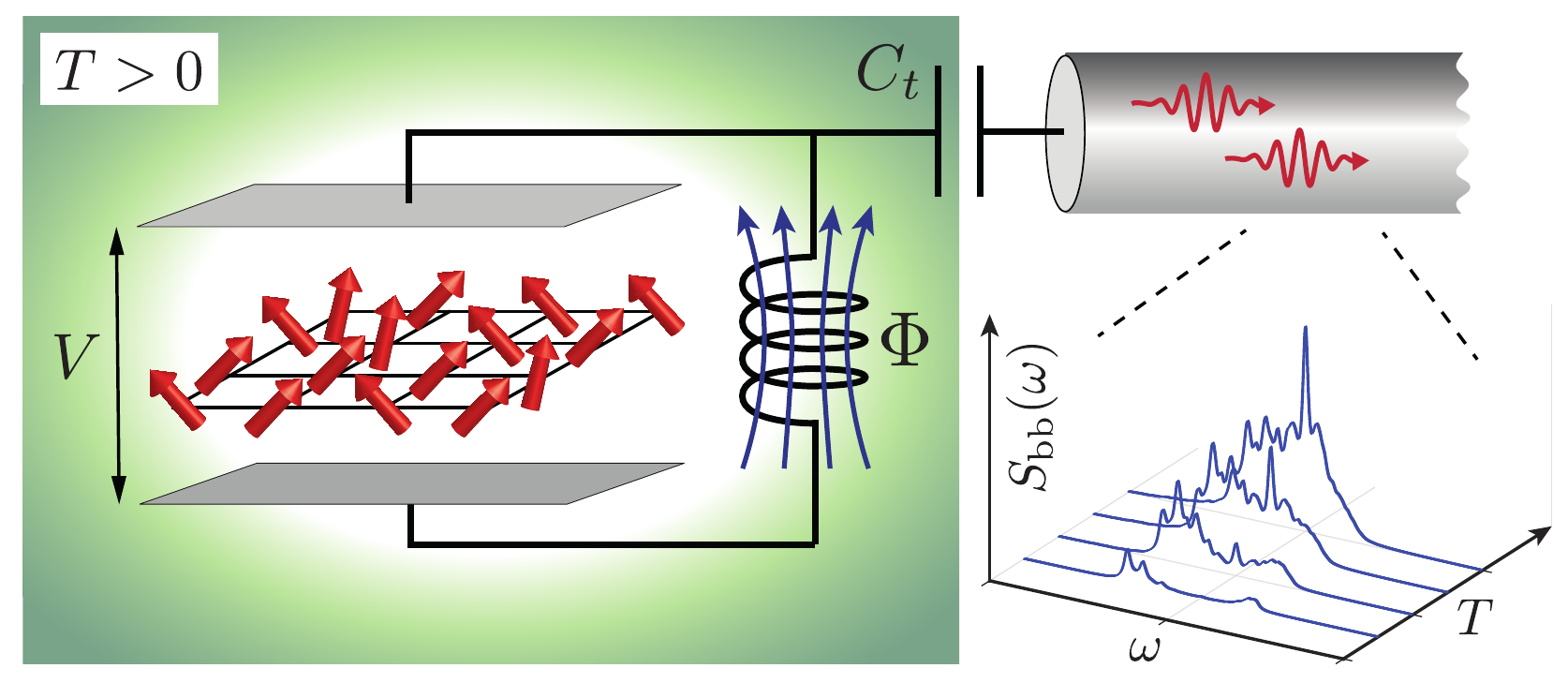}
	\caption{Sketch of a cavity QED setup where an ensemble of dipoles is coupled to the electric field of a lumped-element $LC$ resonator. The system is in thermal contact with a bath of temperature $T$. The black-body spectrum of the cavity mode, $S_{\rm bb}(\omega)$, can be measured through a weak capacitive link to a cold transmission line. See text for more details.}
	\label{Fig1:Setup}
\end{figure}

In this paper we study the thermodynamics of cavity and circuit QED systems within the framework of the extended Dicke model (EDM)~\cite{Jaako2016, DeBernardis2018}. Although based on several simplifications, such as the two-level and the single-mode approximation, this model remains consistent with basic electrodynamics at arbitrary interaction strengths and distinguishes explicitly between static and dynamical electric fields. It thus allows us to evaluate the free energy of the most relevant dynamical degrees of freedom in cavity QED and to study the thermal equilibrium states of the combined system for a macroscopic number of $N\gg1$ dipoles and in various coupling regimes. 

Our analysis shows, first of all,  that in the collective USC regime, where $G=g\sqrt{N}$ is comparable to the cavity frequency $\omega_c$, but where the coupling $g$ between the cavity and each individual dipole is still small, the coupling-induced corrections to the free energy scales as $F_g\sim \hbar g^2N/\omega_c>0$. This very generic result, which also holds for arbitrary dipolar systems, shows that the coupling to a cavity mode leads to a positive shift of the free energy and when taking the limit $N\rightarrow \infty$ for a fixed value of $G$, the changes in the free energy per particle, $F_g/N$, vanish. Both findings contradict the common intuition built upon the analysis of the Dicke model, which predicts negative corrections to the free energy~\cite{CanaguierDurand2013} and that a large collective coupling to a quantized field mode can induce substantial modifications of material properties~\cite{Hepp1973,Wang1973,Carmichael1973}.
Our results are, however, consistent with similar conclusions obtained in studies about molecular properties in the ground state of cavity QED systems~\cite{Galego2015,Cwik2016,Martinez2018}, and can be intuitively explained by a simple polariton picture: In the collective USC regime the cavity field only couples to a single collective dipole mode while the other $N-1$ orthogonal excitation modes remain unaffected. Therefore, the presence of a single cavity mode should not have a considerable impact on the thermodynamics of a macroscopic ensemble of dipoles.

Surprisingly, in the regime $g\sim \omega_c$ this intuition is no longer true and  we find that the coupling to the cavity can indeed influence quantities such as the electric susceptibility or the specific heat,  or even the phase transition temperature of a ferroelectric material. This creates a highly unusual situation where the addition of a single degree of freedom changes the thermodynamics of a macroscopic system. Further, we show that the different coupling regimes of cavity QED result in very distinct features in the black-body spectrum of the cavity. As $g$ in increased, the spectrum evolves from the usual polariton doublet into a broad and disordered set of lines and, finally, collapses again to a single resonance. At the same time we find that already at moderate coupling strengths, the light-matter interaction can either enhance or suppress the total radiated power. Therefore, the analysis of the EDM already provides many conceptually important predictions, which can serve as a basis for more detailed investigations of thermal effects in real and artificial light-matter systems.

The remainder of the paper is structured as follows: In Sec. \ref{sec:EDM} and Sec. \ref{sec:FincQED} we first introduce the EDM and discuss some general properties of the free energy of a cavity QED system in different coupling regimes. In Sec. \ref{sec:cpara} we then analyze in more detail the cavity-induced modifications for the cases of paraelectric and ferroelectric ensembles of dipoles. Finally, in Sec. \ref{sec:bbr} we evaluate the black-body spectrum of the cavity mode in different coupling regimes and we conclude our work in Sec. \ref{sec:conclusions}. The appendices A-D contain  additional details about different approximation methods for the free energy and the derivation of the emission spectrum.

\section{Model}
\label{sec:EDM}

We consider a generic cavity QED setup, where $N$ two-level dipoles are coupled to a single electromagnetic mode. However, since we are interested in both thermal und USC effects, we can restrict our discussion to cavity and circuit QED setups in the GHz to THz regime, where these effects are experimentally most relevant. In this case the confined electromagnetic field can be represented by the fundamental mode of a lumped-element $LC$ resonator~\cite{DeBernardis2018,Todorov2014} with capacitance $C$ and inductance $L$ (see Fig. \ref{Fig1:Setup}).  This configuration also ensures that all higher EM excitations are well separated in frequency and that the electric field is approximately homogeneous across the ensemble of dipoles. The dipoles themselves are modeled as effective two-level systems with states $|0\rangle$ and $|1\rangle$. The two states are separated by an energy $\hbar \omega_0$ and they are coupled via an electric transition dipole moment $\mu$ to the electric field. 

\subsection{Hamiltonian}

The Hamiltonian of the whole cavity QED system can be written as 
\begin{equation}\label{eq:HemHdip}
H_{\rm cQED}=  H_{\rm em} +  H_{\rm dip},
\end{equation}
where the two terms represent the energies of the EM mode and the dipoles, respectively. We model the bare dynamics of the dipoles by a spin Hamiltonian of the form
 \begin{equation} 
H_{\rm dip} =\frac{\hbar \omega_0}{2} \sum_{i=1}^N\sigma_z^i   + \hbar  \sum_{i,j=1}^N \frac{J_{ij}}{4} \sigma_x^i \sigma_x^j,
\end{equation}
where the $\sigma_k^i$ are the usual Pauli operators for the $i$-th dipole. The couplings $J_{ij}$ account for the effect of static dipole-dipole interactions as well as possible other types of non-electromagnetic couplings between the two-level systems. For all of the explicit calculations below we will consider the special case of all-to-all interactions, $J_{ij}=J/N$.  In this limit the spin Hamiltonian reduces to the Lipkin-Meshkov-Glick (LMG) model~\cite{LMG}
\begin{equation}
H_{\rm dip} =\hbar \omega_0 S_z   + \frac{\hbar J}{N} S_x^2\equiv H_{\rm LMG}, 
\end{equation}
where the $S_k=1/2\sum_i \sigma_k^i$ are collective spin operators. For the current purpose, this model is sufficient to capture the qualitative features of non-interacting $(J=0)$, ferroelectric $(J<0)$ and anti-ferroelectric $(J>0)$ dipolar systems, while still being simple enough to allow for exact numerical simulations for moderate system sizes. However, we emphasize that none of the general conclusions and theoretical approaches in this work depend on the assumption of all-to-all interactions and can be extended to arbitrary dipolar systems using more sophisticated numerical techniques~\cite{Schuler2020}.

In the lumped-element limit, the energy of the EM mode is given by 
\begin{equation} \label{eq:HEM}
H_{\rm em}= \frac{CV^2}{2} + \frac{\Phi^2}{2L}  =  \frac{(Q+\mathcal{P}/d)^2}{2C} + \frac{\Phi^2}{2L},
\end{equation}
where $V$ is the voltage difference across the capacitor plates and $\Phi$ the magnetic flux. After the second equality sign we have expressed the capacitive energy in terms of the total charge $Q$, which is the variable conjugate to $\Phi$ and obeys $[\Phi,Q]=i\hbar$ in the quantized theory. For a sufficiently homogeneous field, the charge is given by $Q=CV-\mathcal{P}/d$, where $\mathcal{P} =\sum_i \mu \sigma_x^i$ is the total polarization and $d$ is the distance between the capacitor plates. As usual we express $\Phi$ and $Q$ in terms of annihilation and creation operators $a$ and $a^\dag$ as
\begin{equation}
Q=Q_0(a+a^\dag),\qquad   \Phi =i \Phi_0 (a^\dag -a) , 
\end{equation}
where $Q_0=\sqrt{\hbar/(2 Z)}$, $\Phi_0=\sqrt{\hbar Z/2}$ and $Z=\sqrt{L/C}$ is the cavity impedance.  Altogether, we obtain the canonical cavity QED Hamiltonian~\cite{DeBernardis2018} 
\begin{equation}\label{eq:HcQED}
\begin{split}
H_{\rm cQED}=  \hbar \omega_c a^\dag a  + \hbar g(a+a^\dag)S_x +\frac{\hbar g^2}{\omega_c}S_x^2 + H_{\rm dip},
\end{split}
\end{equation}
where $\omega_c=1/\sqrt{LC}$ and $g=\mu Q_0 / (\hbar C d)$ is the coupling strength. 

The form of $H_{\rm cQED}$ given in Eq.~\eqref{eq:HcQED} allows a clear distinction between electrostatic and dynamical effects. Here the terms $\sim J_{ij}\sigma_x^i\sigma_x^j$ represent the electrostatic energy of the ensemble with a fixed orientation of the dipoles. This energy might be modified in the presence of metallic cavity mirrors~\cite{DeBernardis2018,Vukic2012,Schuler2020}, but it is independent of the frequency or the vacuum field amplitude of the dynamical mode. The coupling to the dynamical field is then proportional to $g$ and includes the collective dipole-field coupling as well as the so-called $P^2$-term $\sim S_x^2$~\cite{Jaako2016,DeBernardis2018,Todorov2014}. This distinction shows that the regular Dicke model, which is recovered for $J_{ij}=-g^2/\omega_c$, describes a very special case of a dipolar system with attractive all-to-all dipole-dipole interactions. Although such a scenario can be realized in circuit QED~\cite{Bamba2016}, the analysis of this specific model does not provide much insights on the behavior of more general cavity QED systems. 

\subsection{Observables}
Apart from $H_{\rm cQED}$, which determines the dynamics and the equilibrium states of the system, it is also important to identify the relevant measurable observables. For the dipoles, quantities of interest are the population imbalance, $\langle S_z\rangle$,  or the polarization along the cavity field, $\langle S_x\rangle\sim \langle \mathcal{P}\rangle$, etc.  Since the operator $Q$ for the total charge includes the induced charges from the dipoles, its value is typically not directly measurable. Therefore, the relevant observables for the cavity mode are the magnetic flux $\Phi$ and the voltage drop $V$ (see Fig.~\ref{Fig1:Setup}) and it is convenient to introduce the displaced photon annihilation operator
 \begin{equation}
 A = a+\frac{g}{\omega_c} S_x, \qquad [A,A^\dag ]=1.
 \end{equation}
With the help of this definition we obtain~\cite{DeBernardis2018,Settineri2019}
 \begin{equation}
 V=V_0 (A+A^\dag), \qquad \Phi=i\Phi_0 (A^\dag -A),
 \end{equation}
 where $V_0=Q_0/C$, and the total Hamiltonian can be written in a compact form as
\begin{equation}
H_{\rm cQED}=    \hbar \omega_c A^\dag A+ H_{\rm dip}.
\end{equation}
By comparing with Eq.~\eqref{eq:HemHdip}, we see that the expectation value of $\langle A^\dag A\rangle$ can be interpreted as the energy of the dynamical cavity mode in units of $\hbar \omega_c$. This is in contrast to the conventional photon number $\langle a^\dag a\rangle$, which depends on the chosen gauge~\cite{DeBernardis2018,Settineri2019} and has no simple interpretation in a strongly coupled cavity QED system. Note, however, while $A+A^\dag$, $A^\dag A$, etc. represent physical properties of the cavity mode only, the operators $A$ and $A^\dag$ do not commute with all the dipole operators and on a formal level we must still use $a$ and $a^\dag$ to represent the independent cavity degree of freedom.

While we focus here on a lumped-element realization of the EM mode as an explicit example,
the model in Eq. (\ref{eq:HEM}) and all the results discussed in this work can be readily applied to arbitrary cavity QED systems using the replacements~\cite{DeBernardis2018,Keeling2007,Vukic2012,Todorov2014,Settineri2019,PhotonsAndAtoms} 
\begin{equation}
V\rightarrow E, \qquad Q\rightarrow D, \qquad \Phi\rightarrow B.
\end{equation}
Here $E$, $D$ and $B$ are the operators for the electric field, the displacement field and the magnetic field, respectively. For a detailed derivation and justification of Hamiltonian~\eqref{eq:HcQED} in dipolar cavity QED and circuit QED settings, see Refs.~\cite{Jaako2016,DeBernardis2018,DeBernardis2018b}.

\section{The free energy in cavity QED}
\label{sec:FincQED}

By assuming that the cavity QED system is weakly coupled to a large reservoir of temperature $T$, the resulting equilibrium state of the system is
\begin{equation}
\rho_{\rm th} =  \frac{1}{Z} e^{-\beta H_{\rm cQED}},   
\end{equation}
where $\beta=1/(k_BT)$ and $Z={\rm Tr}\{e^{-\beta H_{\rm cQED}}\}$ is the partition function. In this case the central quantity of interest is the free energy,
\begin{equation}
F= -k_BT \ln (Z)= F_c^0 + F_{\rm dip}^0+F_g,
\end{equation}
which we divide into the free energies $F_c^0$ and $F_{\rm dip}^0$ of the decoupled subsystems and a remaining contribution $F_g$. In the following discussion we will mainly focus on the coupling-induced part of the free energy, $F_g$, which allows us to separate the influence of light-matter interactions from the thermodynamic properties of the bare subsystems.  

For small and moderately large ensembles the partition function $Z$ and the resulting free energy can be evaluated by diagonalizing $H_{\rm cQED}$ numerically. For the LMG model, which conserves the total spin $s$, this can be done for each spin sector separately and we obtain
\begin{equation}\label{eq:Z} 
Z= \sum_{s} \zeta_{s,N}  Z_s.
\end{equation}
Here $Z_s$ is the partition function of $H_{\rm cQED}$ constrained to a total spin $s$ and~\cite{Hepp1973}
\begin{equation}\label{eq:multiplicity}
\zeta_{s,N} = \frac{N! (2s+1)} {(\frac{N}{2}-s)! (\frac{N}{2} + s + 1)! }
\end{equation}
accounts for the multiplicity of the respective multiplet due to permutation symmetry. For small and moderate temperatures, we use this approach to evaluate the exact free energy for systems with $N\approx 1-100$ dipoles. More details about the numerical calculations are given in Appendix~\ref{app:Numerics}.

\subsection{Mean-field theory} 
In the analysis of the regular Dicke model~\cite{Hepp1973,Wang1973,Carmichael1973} with $N\gg1$, a frequently applied approximation for evaluating the free energy is based on the mean-field decoupling of the dipole-field interaction, 
\begin{equation}\label{eq:MFdecoupling}
(a+a^\dag)S_x \rightarrow (\alpha+\alpha^*) S_x + (a+a^\dag) \Sigma_x - (\alpha+\alpha^*) \Sigma_x,
\end{equation}
where the expectation values $\alpha=\langle a\rangle$ and $\Sigma_x=\langle S_x\rangle$ must be determined self-consistently. Under this mean-field approximation Hamiltonian~\eqref{eq:HcQED} can be written as the sum of two independent parts,
\begin{equation}
H_{\rm MF}= \hbar \omega_c a^\dag a + \hbar g(a+a^\dag) \Sigma_x + H^{\rm MF}_{\rm dip}(\alpha,\Sigma_x),
\end{equation}
where $H^{\rm MF}_{\rm dip}(\alpha,\Sigma_x)=H_{\rm dip} + \hbar g(\alpha+\alpha^*)(S_x-\Sigma_x) +\hbar g^2S_x^2/\omega_c$. The first two terms describe the energy of a displaced oscillator, which is minimized for $\alpha= -(g/\omega_c) \Sigma_x$. With the help of this relation between $\alpha$ and $\Sigma_x$, the total partition function in mean-field approximation is given by
\begin{equation}\label{eq:ZMF}
Z_{\rm MF}(\Sigma_x) = Z_c^0  \times \bar  Z_{\rm dip}^{\rm MF}(\Sigma_x).
\end{equation}
Here, 
the  first factor is the partition function of the bare cavity and $\bar Z_{\rm dip}^{\rm MF}(\Sigma_x)={\rm Tr}\{\exp[-\beta \bar H^{\rm MF}_{\rm dip}(\Sigma_x)]\}$ is the partition function of an ensemble of dipoles with effective Hamiltonian (which includes the constant energy shift from the displaced oscillator)
\begin{equation}\label{eq:HdipMF}
\bar H^{\rm MF}_{\rm dip}(\Sigma_x)= H_{\rm dip} + \frac{\hbar g^2}{\omega_c}\left( S_x- \Sigma_x\right)^2.
\end{equation}
The free energy for the whole system in mean-field approximation is then given by 
\begin{equation}
F_{\rm MF}={\rm min}_{\Sigma_x}\{  -k_BT \ln \left[Z_{\rm MF}(\Sigma_x)\right]\}, 
\end{equation}
and $F^{\rm MF}_g=F_{\rm MF}-F_c^0-F_{\rm dip}^{0}$ are the corresponding coupling-induced corrections. Note that the minimization of the free energy also ensures that the self-consistency condition $\langle S_x\rangle = \Sigma_x$ is satisfied. 


\begin{figure}
\centering
	\includegraphics[width=\columnwidth]{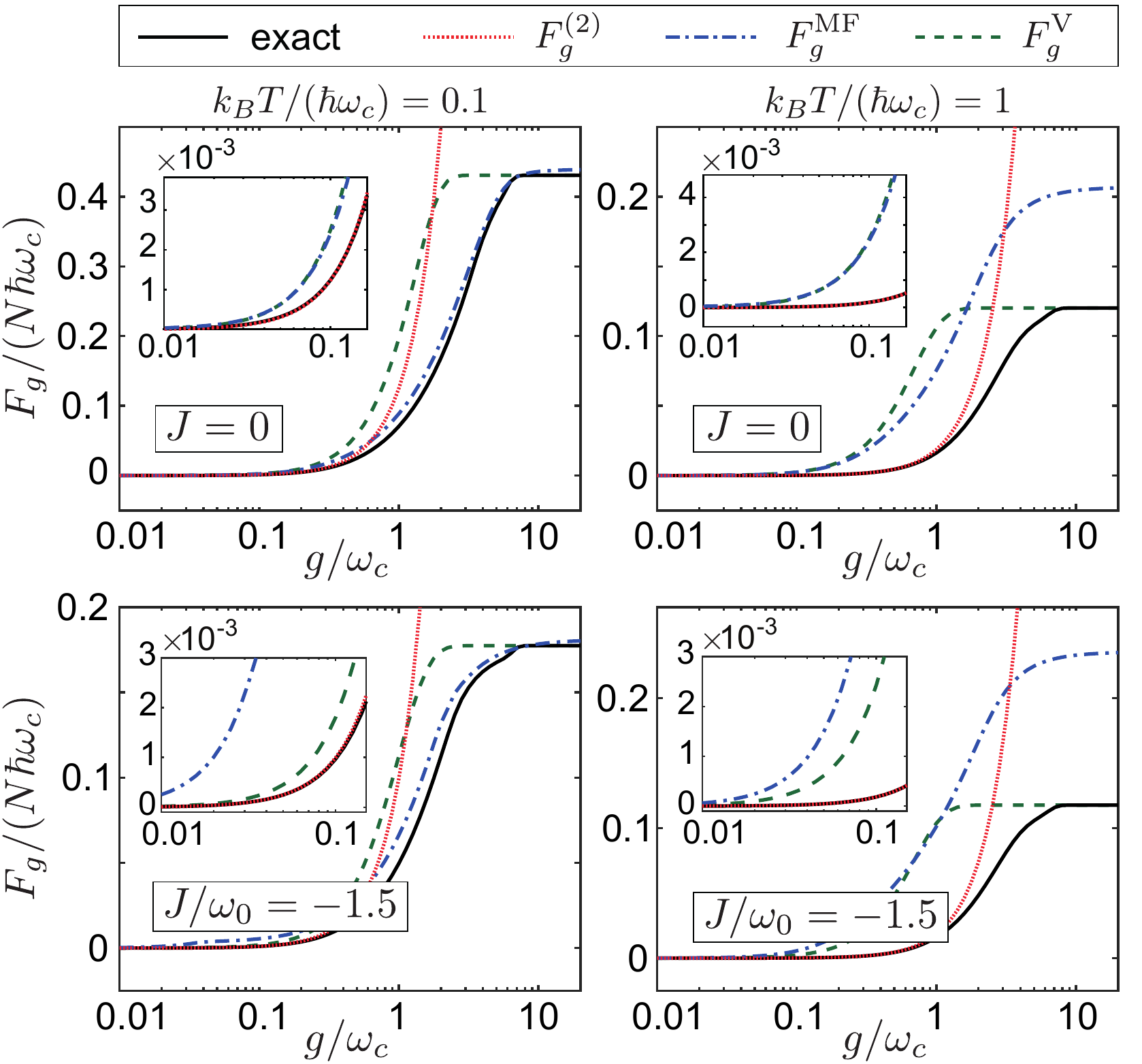}
	\caption{Dependence of the coupling-induced part of the free energy, $F_g$, on the cavity-dipole coupling strength, $g$. This dependence is shown in the individual plots for different temperatures and dipole-dipole coupling strengths, $J$, and for $\omega_c=\omega_0$.  In each plot the exact numerical results for $N=20$ dipoles are compared with  approximate results obtained from mean-field theory ($F_g^{\rm MF}$), second-order perturbation theory ($F_g^{(2)}$) and from a variational calculation ($F_g^{\rm V}$).
	}
	\label{fig:Fg_plot}
\end{figure}

Equation~\eqref{eq:HdipMF} shows that cavity-induced corrections to the thermodynamic properties of a dipolar system are only affected by fluctuations, but not by the mean orientation of the dipoles. Therefore, by applying a second mean-field decoupling for the dipoles (see Appendix~\ref{app:LMG_MF}),  the effect of the cavity vanishes completely and $F^{\rm MF}_g=0$. 
We conclude that a full mean-field treatment, as frequently employed to study the ground states and thermal phases of the Dicke model or of collective spin models~\cite{Das2006,Quan2009,Wilms2012}, cannot be used to analyze the thermodynamics of actual cavity QED systems.

To take fluctuations of the dipoles into account we can evaluate the partition function of the spin system, $\bar Z_{\rm dip}^{\rm MF}(\Sigma_x)$, exactly. 
In this case we find that in the paraelectric phase, where $\Sigma_x=0$, the cavity induces a renormalization of the interaction term, $J\rightarrow J+g^2N/\omega_c$. This renormalization becomes a substantial modification of the dipolar system already in the collective USC regime, $g\sqrt{N}\sim \sqrt{\omega_0\omega_c}$, and could, at first sight, even prevent the ferroelectric instability for $J<0$. While such a shift of the phase transition point is not observed when a proper minimization over $\Sigma_x$ is carried out, a comparison with the exact free energy in Fig.~\ref{fig:Fg_plot} shows that  mean-field theory systematically overestimates the influence of the cavity mode, in particular at higher temperatures. This somewhat counterintuitive trend can be traced back to the fact that  the mean-field decoupling in Eq.~\eqref{eq:MFdecoupling} neglects contributions  which are second order in $\hbar g (a+a^\dag) S_x$ and scale approximately as
\begin{equation}
H_g^{(2)}\sim - \frac{\hbar g^2}{\omega_c} S_x^2.
\end{equation}
Therefore, the mean-field decoupling neglects an essential contribution from the light-matter interaction and the approximation becomes uncontrolled. 

As shown in the examples in Fig.~\ref{fig:Fg_plot}, at larger coupling parameters $g/\omega_c$ and low temperatures, the mean-field predictions agree reasonably well with the exact results. However, this agreement seems to be accidental since at higher temperatures there are again substantial deviations and the limit $g/\omega_c\gg 1 $ is not reproduced correctly. Although not shown explicitly, a very similar trend is also found in the anti-ferroelectric case, $J>0$. In summary, these results for small and large couplings indicate that also on a qualitative level Hamiltonian $\bar H^{\rm MF}_{\rm dip}(\Sigma_x)$ does not correctly capture the influence of the cavity mode.

\subsection{Collective USC regime}\label{subsec:CollectiveUSC}

Many cavity QED experiments are operated in the regime $G=g\sqrt{N}\lesssim \omega_c$ and $N\gg1$, where the collective coupling $G$ can become comparable to the photon frequency, but the coupling of each individual dipole to the cavity mode is still very small, $g\ll \omega_c$. In this regime, we can treat the dipole-field interaction,
\begin{equation}\label{eq:Hg}
H_g=\hbar  g(a+a^\dag) S_x +  \frac{\hbar g^2}{\omega_c} S_x^2,
\end{equation} 
as a small perturbation and expand the free energy in powers of $g$. As a result of this calculation, which is detailed in Appendix~\ref{app:Perturbation}, we obtain the lowest-order correction to the bare free energy. It can be written in the form 
\begin{equation}
F^{(2)}_g =  N\frac{\hbar g^2}{4\omega_c} f_g.
\end{equation}
The dimensionless function $f_g\sim O(1)$ contains two contributions, one arising from the average value of the $S^2_x$ term and a second-order contribution from the linear coupling term, $ \hbar g(a^\dag+a)S_x$. The resulting expression for $f_g$ still involves non-trivial correlation functions of spin operators, which for interacting dipoles must be evaluated numerically. For non-interacting dipoles this calculation can be carried out analytically and we obtain the explicit result  
\begin{equation}\label{eq:fgfull}
f_g(J=0)= \frac{\omega_0^2 -   \omega_0\omega_c \tanh\left(\frac{\hbar \omega_0}{2k_BT}\right) \coth\left(\frac{\hbar \omega_c}{2k_BT} \right)}{(\omega_0^2-\omega_c^2)}. 
\end{equation}
In Fig.~\ref{fig:Fg_plot}, this prediction from perturbation theory is compared to the exact free energy and we find that the cavity-induced corrections to the free energy are very accurately reproduced by $F_g^{(2)}$ at low and high temperatures, even for collective coupling strengths as large as  $G\approx \omega_c$. 

By taking the limit $T\rightarrow 0$, Eq.~\eqref{eq:fgfull} provides us directly with the lowest-order correction to the ground state energy of a cavity QED system~\cite{Ciuti2005},
\begin{equation}\label{eq:E0}
E_0^{(2)}= F^{(2)}_g(T\rightarrow 0,J=0) =  N\frac{\hbar g^2}{4 \omega_c}\frac{\omega_0}{\omega_0+\omega_c},
\end{equation}
which agrees with Hamiltonian perturbation theory. In the opposite high-temperature limit we obtain 
\begin{equation}
F^{(2)}_g(T\rightarrow \infty) \simeq N\frac{\hbar^3 g^2 \omega_0^2 }{48 \omega_c k_B^2T^2}.
\end{equation}
Therefore, the cavity-induced corrections to the free energy vanish quadratically with increasing temperature.
Importantly, we find that for all temperatures $F^{(2)}_g\geq 0$, which is in stark contrast to the negative correction terms obtained within the framework of the regular Dicke model~\cite{CanaguierDurand2013}. More generally, one can show that also for an arbitrary system of interacting two-level dipoles (see Appendix~\ref{app:Perturbation})
\begin{equation}\label{eq:fg2Bound} 
0\leq f_g < \frac{ 4 (\Delta S_x)^2  }{N},
\end{equation} 
where $(\Delta S_x)^2= \langle S_x^2\rangle_0-\langle S_x\rangle^2_0$.  For a regular dipolar system away from a critical point, the spatial extent of the individual correlations, $\langle \sigma_x^i \sigma_x^j\rangle -  \langle \sigma_x^i \rangle \langle \sigma_x^j\rangle$, is finite and the collective fluctuations scale as $ (\Delta S_x)^2 \sim  N$. Therefore, under very generic conditions, by taking the limit $N\rightarrow \infty$ with $G$ kept fixed, we obtain
\begin{equation}\
\lim_{N\rightarrow \infty} \frac{F_g^{(2)}}{N} =0.
\end{equation}
This result confirms our basic intuition that the coupling of many dipoles to a single mode should not affect extensive thermodynamic properties. Note that this  conclusion does not necessarily hold for a stronger scaling of fluctuations, i. e. $(\Delta S_x)^2\sim N^2$. This scaling can be found, for example, in the LMG model for $J<\omega_0$ when the symmetry of the ferroelectric phase is not explicitly broken. However, even in this special case our exact numerical calculations confirm that the bound $f_g<1$ still holds.

\subsection{Non-perturbative regime}
\label{sec:npregime}
The physics of cavity QED changes drastically once $g\sim \omega_c$ and the light-matter interactions become non-perturbative at the level of individual dipoles. To analyze this regime, it is usually more convenient to change to a polaron frame, $\tilde H_{\rm cQED}= UH_{\rm cQED}U^\dag$, via the unitary transformation $U= e^{\frac{g}{\omega_c} S_x ( a^\dagger - a) }$~\cite{Irish2007,Chen2008,LeBoite2020}. In this frame the cavity QED Hamiltonian can be written as \cite{Jaako2016,DeBernardis2018,Stefano2019,Alcalde2012}
\begin{equation}
\tilde H_{\rm cQED}= \hbar \omega_c a^\dag a+ H_{\rm dip} + H_{\rm int},
\end{equation}
where the interaction part now takes the form
\begin{equation}\label{eq:Hint}
H_{\rm int}= \hbar \omega_0 \left(U S_z U^\dag - S_z\right).
\end{equation}
An immediate benefit of the polaron representation is that the interaction is proportional to $\omega_0$. This shows that for $\omega_0\rightarrow 0$ the coupling to the dynamical mode vanishes and we recover the electrostatic limit, $\tilde H_{\rm cQED}(\omega_0\rightarrow 0)=  \hbar \omega_c a^\dag a+ \sum_{i,j} \frac{\hbar J_{ij}}{4} \sigma_x^i\sigma_x^j$.  

For finite $\omega_0$ the effects of $H_{\rm int}$ are more involved. For $T=J=0$ it can be shown that up to second order in $H_{\rm int}$ and for $g/\omega_c\gtrsim 2$, the low-energy behavior of the dipolar system is well-described by the effective spin Hamiltonian~\cite{Jaako2016}
\begin{equation} \label{eq:HSgcorr}
H_{\rm eff} = \hbar \omega_0 \left( e^{-\frac{g^2}{2\omega_c^2} } -1 \right)S_z +  \frac{\hbar \omega_0^2 \omega_c}{2g^2} \left( S_x^2- \vec S^2\right).
\end{equation}
This effective model captures two important signatures of non-perturbative light-matter interactions, which will be relevant for the discussions below.  Firstly, there is a strong suppression of the dipole oscillation frequency when $g/\omega_c\gtrsim 1$.
Secondly, the cavity mediates an all-to-all anti-ferroelectric coupling $\sim \omega_0^2$.  

In principle, we can again apply perturbation theory to evaluate $F_g$ up to second order in $H_{\rm int}$ and extend the results from Sec.~\ref{subsec:CollectiveUSC} into the strong-coupling regime. However, such an approach is only reliable when $\omega_0\ll \omega_c$ and the resulting expressions are much more involved. Therefore, this method is only briefly summarized in Appendix~\ref{app:Perturbation}. As a less accurate, but more intuitive approach we can use the variational principle of Bogoliubov to derive an upper bound $F_{V}$ for the free energy,
\begin{equation} \label{eq:var_principle}
F \leq F^* + \braket{\tilde H_{\rm cQED} - H^*}_{\rho^*} \equiv  F_{V}.
\end{equation}
Here $\rho^*$ is the thermal state and $F^*$ the corresponding free energy for the trial Hamiltonian $H^*$. Based on the discussion above we choose 
\begin{equation}\label{eq:Hstar}
H^*= \hbar \omega_c a^\dag a + \hbar \tilde \omega_0 S_z + \frac{\hbar J}{N}S_x^2,
 \end{equation}
which describes a non-interacting cavity QED system, but with a variable frequency $\tilde \omega_0$. By minimizing $F_{V}$ with respect to $\tilde \omega_0$ for each $g$ we obtain (see Appendix~\ref{app:Variational})
\begin{equation} \label{eq:w0_var}
\tilde \omega_0(g)= \omega_0 e^{-\frac{g^2}{2\omega_c^2}(1 + 2 N_{\rm th})},
\end{equation}
where $N_{\rm th}=1/(e^{\beta \hbar \omega_c}-1)$. While from the comparison in Fig. \ref{fig:Fg_plot} we see that overall $F_g^V=F_V-F_c^0-F_{\rm dip}^0$ does not reproduce the quantitative behavior of $F_g$ very accurately, we will see in the following that there are still many cavity-induced modifications that can be directly explained by this simple renormalization of the dipole frequency.

\section{Para- and ferroelectricity in the USC regime}
\label{sec:cpara}

While the free energy contains all the relevant information about the cavity QED system, we are usually interested in derivative quantities such the susceptibility, the specific heat, etc., or the existence of different phases and the transitions between them. To understand in which way the coupling to a quantized cavity mode can influence such quantities, we discuss in this section three elementary examples. 

\subsection{USC modifications of the Curie law}
\begin{figure}
\centering
	\includegraphics[width=\columnwidth]{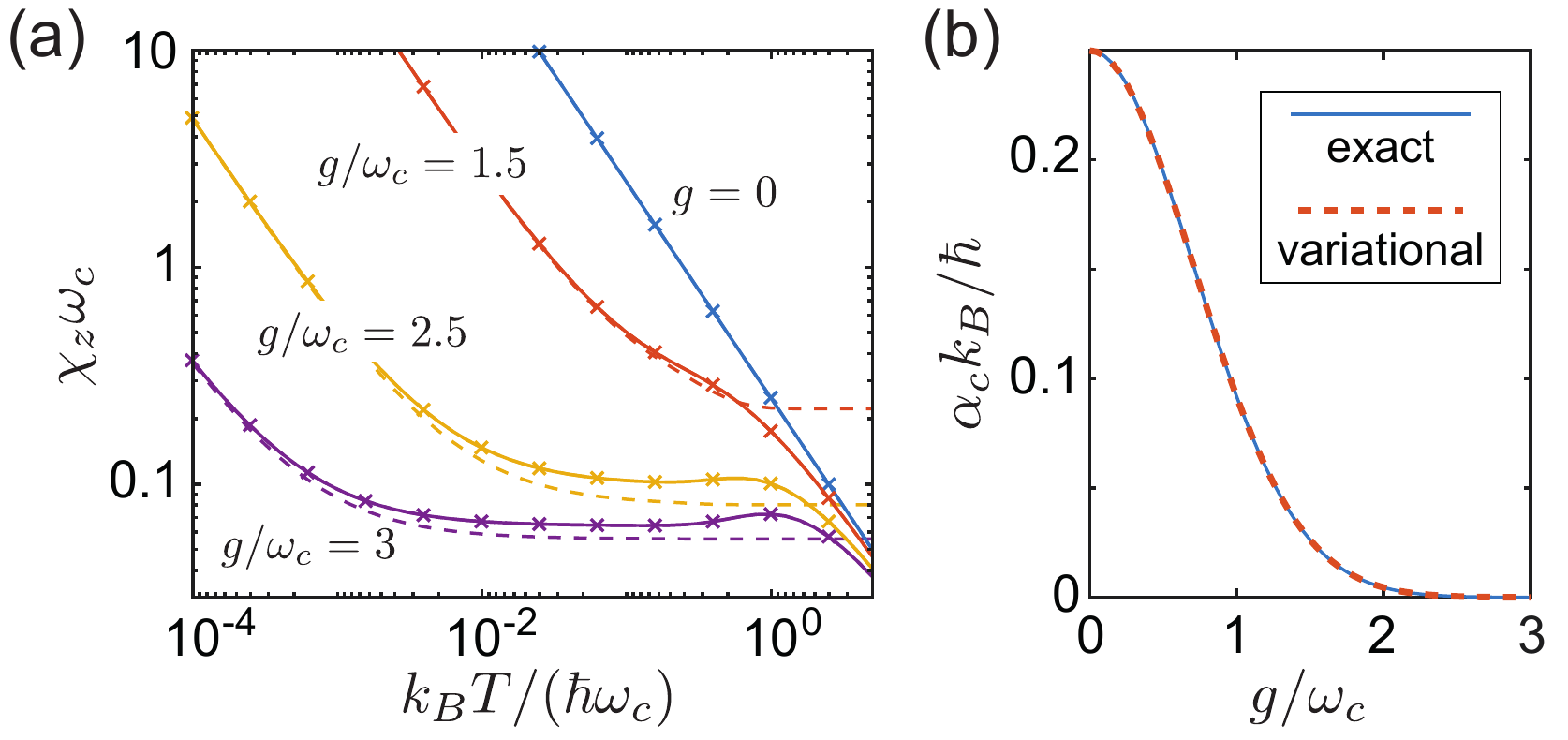}
	\caption{(a) Plot of the zero-field susceptibility $\chi_z$ (solid lines) for different coupling parameters $g/\omega_c$. The dashed lines indicate the predictions from the approximate formula given in Eqs.~\eqref{eq:Chiz_terms}-\eqref{eq:chizoffset}.  The x-markers show the results obtained from the perturbation theory discussed in Appendix \ref{App:Perturbation_w0}.  (b) Dependence of the Curie constant $\alpha_C(g)$ on the dipole-field coupling strength. The exact numerical results are in perfect agreement with the analytic scaling derived in Eq.~\eqref{eq:Curie} from the variational free energy, $F_V$. For all plots $N=20$, $\omega_c=\omega_0$ and $J=0$ have been assumed.}
	\label{fig:Chiz}
\end{figure}

As a first example it is instructive to consider the simplest case of non-interacting dipoles, where
\begin{equation}
H_{\rm dip}= \hbar \omega_0 S_z.
\end{equation}
This means, that in the absence of the cavity, the dipoles form an ideal paraelectric material. For this system, we are interested in the dependence of the population imbalance $ \langle S_z\rangle$ on the level splitting $\omega_0$. Specifically, we consider the limit $\omega_0\rightarrow 0$, where we obtain the zero-field susceptibility
\begin{equation}\label{eq:chiz}
\chi_z  = \left.\frac{1}{N} \frac{\partial \langle S_z\rangle}{\partial \omega_0} \right|_{\omega_0=0}  = -\frac{1}{N \hbar} \left. \frac{\partial^2 F}{\partial \omega_0^2}\right|_{\omega_0=0}
\end{equation}
from the second derivative of the free energy. In the limit of a vanishing bias field the susceptibility follows the usual Curie law
\begin{equation}
\left.\chi_z\right|_{g\rightarrow 0}(T) = \frac{\alpha_C}{T},
\end{equation}
with a Curie constant $\alpha_C=\hbar /(4k_B)$. In the context of cavity QED, the behavior of this quantity for finite $g$ is of particular importance. Since the dipoles decouple from the cavity mode for $\omega_0=0$, the zero-field susceptibility captures the lowest-order deviations from the electrostatic limit. 

In Fig.~\ref{fig:Chiz}(a) we plot $\chi_z(T)$ for a cavity QED system with $J=0$, $N=20$ and different coupling strengths $g$. For small $g$ we still recover the $1/T$ behavior with a small reduction of the Curie constant. In the non-perturbative regime, $g/\omega_c \gtrsim 1$, the modifications become more significant. Although in this regime the susceptibility still diverges for $T\rightarrow 0$, there appears an additional plateau for an intermediate range of temperatures, $T<\hbar \omega_c/k_B$. To understand this behavior we approximate the susceptibility by the two dominant contributions, 
\begin{equation} \label{eq:Chiz_terms}
\chi_z \approx  \frac{\alpha_C(g)}{T} - \frac{1}{N \hbar} \sum_n p_n \left. \frac{\partial^2  E_n}{\partial \omega_0^2}\right|_{\omega_0=0} ,
%
\end{equation}
where $p_n$ and $E_n$ are the thermal occupation probabilities and the energy of the $n$-th eigenstate, respectively.  
The first term emerges from the change of $\langle S_z\rangle$ due to small changes of the thermal populations when $\omega_0$ is varied. Since we are interested in the limit $\omega_0\rightarrow 0$, this change results in the same high-temperature scaling $\sim1/T$ as in the case of free dipoles. This effect is already captured by the variational free energy $F_V$ discussed in Sec. \ref{sec:npregime}, from which we obtain 
\begin{equation}\label{eq:Curie}
\alpha_C(g) \simeq  \frac{\hbar}{4 k_B} e^{-\frac{g^2}{\omega_c^2}(1+2N_{\rm th})}.
\end{equation}
In Fig. \ref{fig:Chiz}(b) we show that this analytic result is in perfect agreement with the low-temperature limit of $\chi_z$ obtained from exact numerical simulations.  

The second term in Eq.~\eqref{eq:Chiz_terms} is the contribution to the susceptibility, which arises from quadratic changes of the energy eigenstates with varying $\omega_0$. For free dipoles, $H_{\rm dip}\sim \omega_0$ and therefore this contribution is absent  in regular paramagnetic and paraelectric systems.  However, as evident from the effective spin model in Eq.~\eqref{eq:HSgcorr}, for couplings $g>\omega_c$ the energy levels show indeed a quadratic scaling, $E_n \sim \omega_0^2$. From this effective model and by assuming that all spin levels are equally populated, $p_n\approx 1/2^N$, we obtain the approximate result 
\begin{equation}\label{eq:chizoffset}
- \frac{1}{N \hbar} \sum_n p_n  \left. \frac{\partial^2  E_n}{\partial \omega_0^2}\right|_{\omega_0=0} \approx  \frac{\omega_c}{2g^2}.
\end{equation}
As shown in Fig.~\ref{fig:Chiz}(a), this estimate is in very good agreement with the value of the plateau of $\chi_z(T)$ found in exact numerical simulations. For $T\gtrsim \hbar \omega_c/k_B$, the thermal population of the photon states is no longer negligible and the approximate model in Eq.~\eqref{eq:HSgcorr} breaks down. Beyond this point, which for large $g$ is indicated by a small bump, the regular Curie law is approximately recovered. 

Note that since the susceptibility $\chi_z$ is evaluated at $\omega_0=0$, it can be calculated exactly from a second-order perturbation theory in $H_{\rm int}\sim \omega_0$ in the polaron representation (see  Appendix~\ref{app:Perturbation}). Although the resulting expressions must still be evaluated numerically, this method allows us to obtain $\chi_z$ from correlation functions of the dipolar system only. Therefore, this approach can be very useful for performing similar calculations for more complicated dipolar systems.

\subsection{USC modifications of the specific heat}

\begin{figure}
\centering
	\includegraphics[width=\columnwidth]{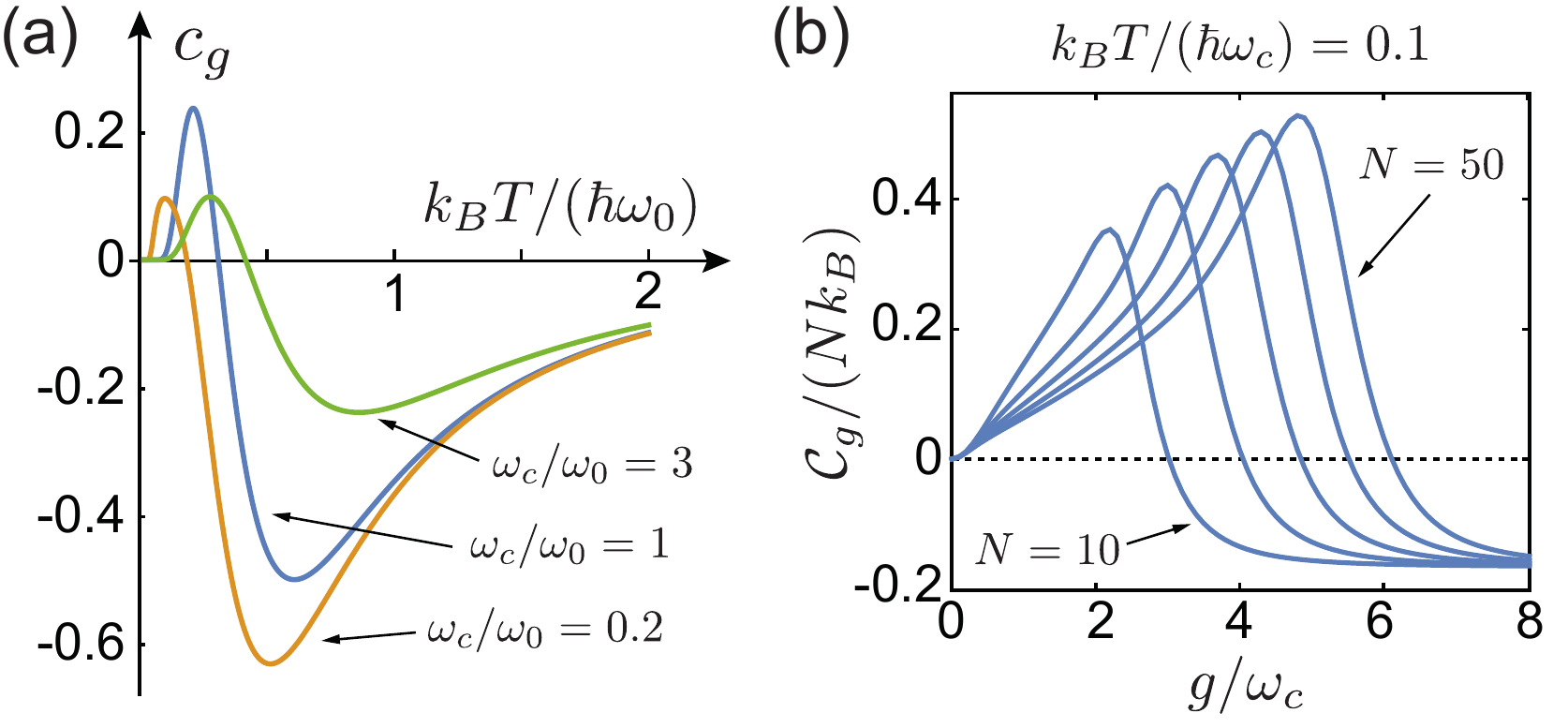}
	\caption{(a) Plot of the dimensionless quantity $c_g$ defined in Eq.~\eqref{eq:Cg}, which determines the corrections to the specific heat, $\mathcal{C}_g/N$, in the collective USC regime. (b) Dependence  of $\mathcal{C}_g$ on the coupling strength $g$ for a fixed temperature and an increasing number of dipoles, $N=10,20,30,40,50$. For this plot $\omega_0 = 0.5 \omega_c$ and a photon cutoff number of $N_{\rm ph} = 100$ has been used.}
	\label{fig:Cg_plot}
\end{figure}

A second quantity of general interest in statistical physics is the heat capacity, 
\begin{equation} \label{eq:C_decomp}
\mathcal{C} = - T \frac{\partial^2 F}{\partial T^2} = \mathcal{C}^0_c+ \mathcal{C}^0_{\rm dip}+\mathcal{C}_g.
\end{equation} 
For a decoupled system, the heat capacities of the cavity and the dipoles, $ \mathcal{C}^0_c$ and $\mathcal{C}^0_{\rm dip}$, are both bounded and scale as $\mathcal{C}^0_c\sim k_B$ and $\mathcal{C}^0_{\rm dip}\sim Nk_B$, respectively. This scaling suggests that for a large ensemble of dipoles and similar energy scales, $\omega_c\approx \omega_0$, the presence of a single cavity degree of freedom should have a negligible contribution to the specific heat $\mathcal{C}/N$ of the combined system.  This can be shown explicitly in the collective USC regime, where for non-interacting dipoles 
\begin{equation}\label{eq:Cg}
\frac{\mathcal{C}_g}{k_BN} \simeq  \frac{\hbar g^2}{4\omega_c(k_BT) } \times c_g(T,\omega_c,\omega_0).
\end{equation}
Here, $c_g= - (k_BT)^2 \partial^2 f_g/\partial (k_BT)^2$ is a dimensionless function, which is independent of $N$ and which is plotted in Fig.~\ref{fig:Cg_plot}(a) for different frequency ratios, $\omega_0/\omega_c$. Therefore, taking the limit $N\rightarrow \infty$ for a fixed $G$, the corrections to the specific heat vanish.

In Fig.~\ref{fig:Cg_plot}(b) we plot $\mathcal{C}_g/N$ for a cavity QED system with an increasing number of $N$ non-interacting dipoles. For small couplings, $G/\omega_c\lesssim 0.5$, the correction is accurately reproduced by the analytic weak-coupling result given in Eq.~\eqref{eq:Cg}. In the non-perturbative regime we observe  substantial modifications. On a qualitative level, these corrections can be understood from a cavity-induced suppression of $\omega_0$, but overall we find that the dependence of $\mathcal{C}_g$ is not very accurately reproduced by the variational ansatz in Eq.~\eqref{eq:Hstar} or any of the other approximation schemes. However, from the exact numerical results plotted in Fig.~\ref{fig:Cg_plot}(b) we see that the maximal correction to the specific heat is 
\begin{equation}
\frac{ |\mathcal{C}_g |}{k_B N} \sim O(1),
\end{equation}
and shifts, but does not vanish with increasing $N$. 
Combined with the behavior of the susceptibility discussed above, this finding demonstrates that in the non-perturbative regime the coupling to a single dynamical field degree of freedom can have a substantial influence on extensive thermodynamic quantities of a large ensemble of dipoles.

\subsection{USC modifications of the ferroelectric phase transition}
\label{sec:cferro}
\begin{figure} [t!] 
\centering
	\includegraphics[width=\columnwidth]{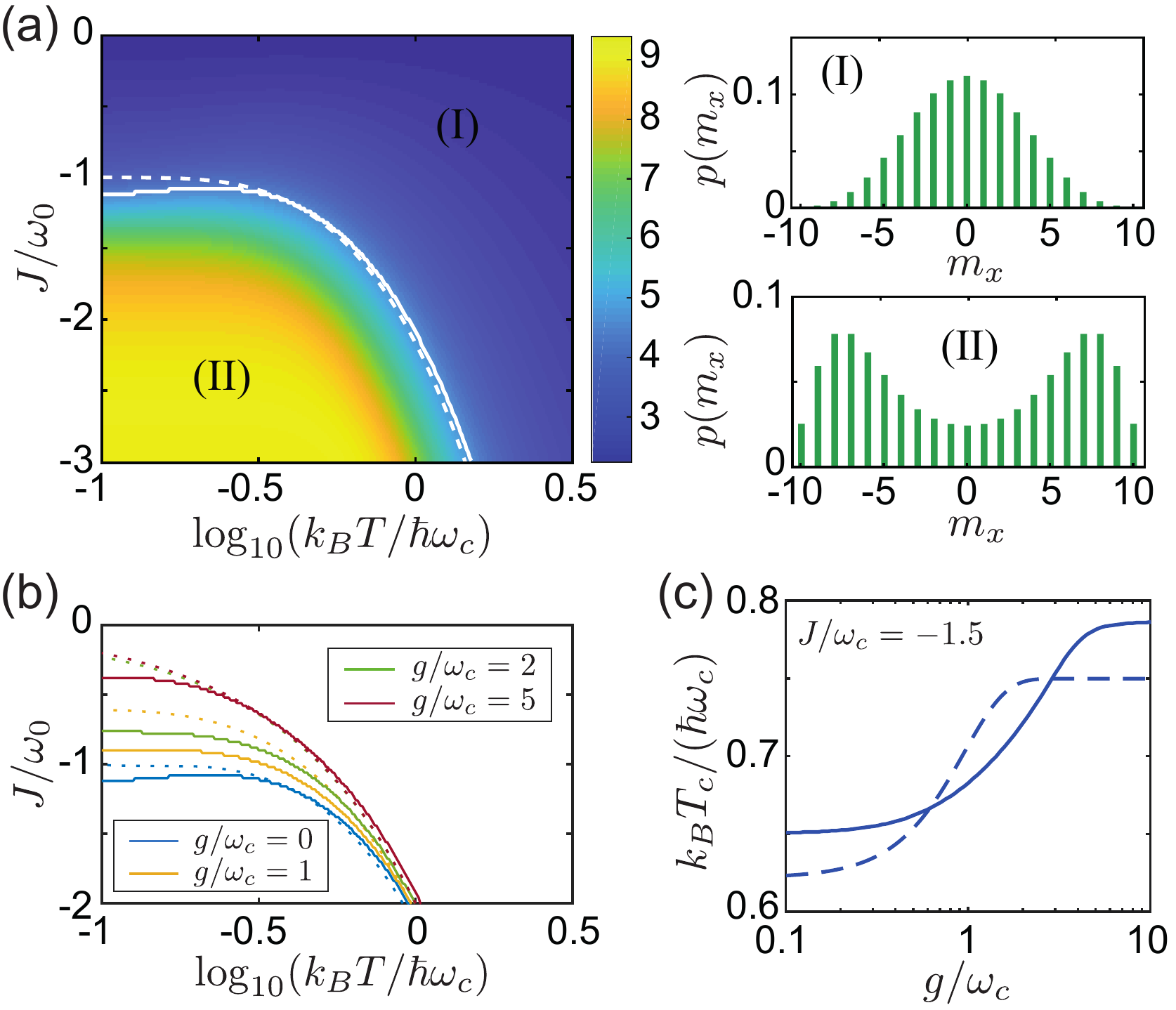}
	\caption{(a) Phase diagram of the LMG model without cavity, where the color scale shows the value of the parameter $\bar m=\sqrt{\langle S_x^2\rangle}$~\cite{Wilms2012} for $N=20$ dipoles. For each point, we also evaluate the probability distribution $p(m_x)$ for the projection quantum number $m_x$, which exhibits a single maximum in the paraelectric phase (I) and two maxima in the ferroelectric phase (II). The transition between the single- and bi-modal distribution is indicated by the solid line, while the dotted line depicts the phase boundary obtained from mean-field theory, see Eq.~\eqref{eq:TcJc}. The same boundaries are shown in (b) for different coupling strengths $g$, where for the mean-field results $\omega_0$ has been replaced by $\tilde \omega_0$. (c) Dependence of the critical temperature $T_c$ on the coupling parameter $g/\omega_c$ for a fixed inter-dipole coupling strength of $J/\omega_c = -1.5$. In all plots $\omega_0=\omega_c$ and $N=20$.}
	\label{fig:ferroPT}
\end{figure}

A central topic of interest in the field of USC cavity QED is the so-called superradiant phase transition, which is predicted for the ground and thermal equilibrium states of the standard Dicke model~\cite{Hepp1973,Wang1973,Carmichael1973}. While in more accurate models for light-matter interactions this transition does not occur for non-interacting dipoles~\cite{Rzazewski1975,ViehmannPRL2011,Todorov2012,Bamba2014,Jaako2016,DeBernardis2018,Rokaj2018,Andolina2019},  the system can still undergo a regular ferroelectric phase transition in the case of  attractive electrostatic interactions, $J_{ij}< 0$. Within the framework of the LMG model, such a transition is well-described by a mean-field decoupling of the collective interaction term, $S_x^2\rightarrow 2 \Sigma S_x - \Sigma^2$ (see Appendix \ref{app:LMG_MF}), from which one can derive the general relation between the critical coupling strength $J_c$ and the critical temperature $T_c$~\cite{Das2006,Quan2009,Wilms2012},
\begin{equation}\label{eq:TcJc}
 \tanh\left(  \frac{\hbar \omega_0 }{2k_BT_c}\right) = -\frac{\omega_0}{J_c}. 
\end{equation}
For $T\rightarrow 0$ there is a critical coupling strength $J_c= - \omega_0$, beyond which the dipoles enter a ferroelectric phase with $\langle S_x\rangle \neq 0$. For $\omega_0\rightarrow 0$ this phase only exists below a critical temperature $T_c=-\hbar J/(2k_B)$, which is just the transition temperature of the classical Ising model. For arbitrary $\omega_0$, the phase boundary of the LMG model in the limit $N\rightarrow \infty$ is indicated by the dashed line in Fig.~\ref{fig:ferroPT}(a).

Since symmetry breaking does not occur for finite $N$, the criterion $\langle S_x\rangle\neq 0$ cannot be used to characterize the ferroelectric phase in exact numerical calculations. In Fig.~\ref{fig:ferroPT}(a) we show  instead the quantity $\bar m= \sqrt{\langle S_x^2\rangle}$, which provides a good indicator for the ferroelectric phase of the LMG model~\cite{Wilms2012}. However, for the rather small numbers of dipoles assumed in the simulations of the full cavity QED model below, the variation of $\bar m$ around the phase transition line is still rather smooth.  Therefore, for the following analysis we consider instead the probability distribution $p(m_x) = {\rm Tr}\{ \mathbbm{P}_{m_x} \rho_{\rm th}\} $, where $\mathbbm{P}_{m_x} = \sum_s \mathbb{P}_{s,m_x}$ and $ \mathbb{P}_{s,m_x} $ is the projector on all states with $S_x\ket{\psi} = m_x\ket{\psi}$ and total spin $ s $.  In this case, we can define the phase boundary as the line where this function changes from a single to a bi-modal distribution, as illustrated in Fig.~\ref{fig:ferroPT}(a). For the bare LMG model, this approach reproduces very accurately the phase boundary derived from mean-field theory, even for a small number of $N=20$ dipoles.

When the dipoles are coupled to the cavity mode, a finite polarization $\langle S_x\rangle\neq 0$ is naturally associated with a non-vanishing expectation value of $\langle a\rangle \simeq -g/\omega_c \langle S_x\rangle$, similar to what is expected for the superradiant phase in the Dicke model. Note, however,  that this expectation value is real and corresponds to a finite charge (or displacement field) $\langle Q\rangle \sim \langle a+a^\dag\rangle$.  
The relevant cavity observables, $\langle V\rangle \sim \langle A+A^\dag\rangle $ and  $\langle \Phi\rangle \sim i\langle A^\dag-A\rangle$, are not affected by this transition~\cite{DeBernardis2018,Keeling2007}. For the ground state, it has further been shown that in the collective USC regime, i.e. when $G\sim \sqrt{\omega_0\omega_c}$ but $g\ll \omega_c$, also the transition point is not influenced by the coupling to the dynamical cavity mode~\cite{DeBernardis2018}. This is no longer the case when $g\sim \omega_c$.

In the current study we are primarily interested in USC effects beyond the ground state and show in Fig.~\ref{fig:ferroPT}(b) the coupling-induced modification of the phase boundary in the whole $T-J$ plane for different values of $g/\omega_c$. In this plot, the exact analytic results are compared with a modified mean-field theory, where in Eq.~\eqref{eq:TcJc}  the bare dipole frequency $\omega_0$ is replaced by  the renormalized frequency $\tilde \omega_0$ given in Eq.~\eqref{eq:w0_var}. From this comparison we find that  the variational free energy $F_V$ captures the overall trend, although the actual phase transition line deviates from the exact results, in particular for $g/\omega_c >1$ and for low temperatures. In Fig.~\ref{fig:ferroPT}(c) we fix the value of $J$ and plot the dependence of the critical temperature on the coupling strength $g$. Consistent with the other examples above, we observe only minor corrections for $G\lesssim \omega_c$, but a substantial increase of $T_c$ for couplings $g/\omega_c\gtrsim 1$. This means that in this coupling regime the presence of the cavity mode stabilizes the ferroelectric phase against thermal fluctuations. This behavior is qualitatively reproduced by the modified mean-field ansatz.

\section{Black-body radiation}
\label{sec:bbr}
The emission spectrum of a hot body was one of the first examples that could not be explained by combining the otherwise very successful theories of statistical mechanics and electromagnetism. In the correct quantum statistical derivation of the black-body spectrum it is assumed that the EM field thermalizes through weak interactions with the material, but that  it can  be treated otherwise as a set of independent harmonic modes. Therefore, it is particularly interesting to see how the thermal emission spectrum of a cavity mode changes under strong and USC conditions~\cite{Ridolfo2013,Ridolfo2013c,Chervy2018}. 

\subsection{Power spectral density}

In the setup shown in Fig. \ref{Fig1:Setup}, the black-body spectrum can be measured, for example, by coupling the cavity via a weak capacitive link to a cold transmission line. The emitted power will then be proportional to the fluctuations of the voltage operator $V=V_0(A+A^\dag)$ (see also Ref.~\cite{Settineri2019}). By assuming that the transmission line can be modeled as an Ohmic bath and that the capacitive link is sufficiently weak, we can write the spectrum of the emitted black-body radiation as (see Appendix~\ref{app:Emission})
\begin{equation}\label{eq:Spectrum}
S_{\rm bb}(\omega)=\frac{\hbar \kappa \gamma}{ 2\pi \omega_c }  \sum_{n>m} \frac{e^{-\beta E_n}}{Z} \frac{ \omega_{nm}^2 |\langle E_n| A+A^\dag |E_m\rangle |^2  }{(\omega - \omega_{nm})^2 + \gamma^2/4},
\end{equation}
where $\omega_{nm}=(E_n-E_m)/\hbar$ are the transition frequencies between the eigenstates $|E_n\rangle$ of $H_{\rm cQED}$ with energies $E_n$. In Eq.~\eqref{eq:Spectrum}, $\kappa$ denotes the decay rate of the bare cavity into the transmission line. In addition, we have introduced a phenomenological rate $\gamma$ to account for a small but finite thermalization rate with the surrounding bath. For consistency we require $\kappa\ll\gamma$ and $\gamma\ll |E_n-E_m|/\hbar$, but otherwise the precise values of $\kappa$ and $\gamma$ are not important. 

\begin{figure}
\centering
	\includegraphics[width=\columnwidth]{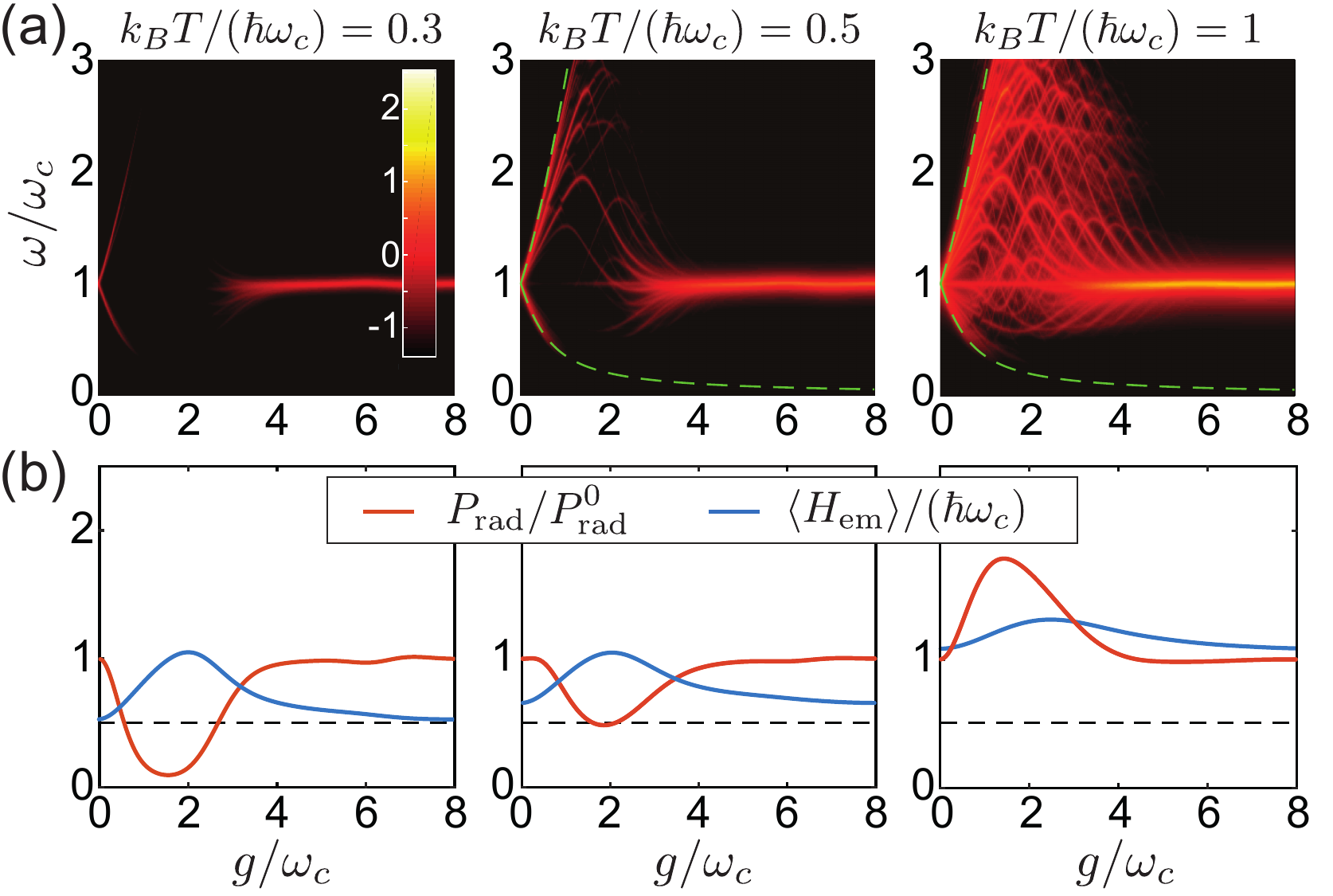}
	\caption{(a) The black-body spectrum $S_{\rm bb}(\omega)/(\hbar\kappa)$ is plotted on a logarithmic scale as a function of $g$  for three different temperatures. The green dashed lines indicate the frequencies $\omega_\pm$ of the two polariton modes obtained from a Holstein-Primakoff approximation. (b) Plot of the total emitted power, $P_{\rm rad}$, and the average value of the EM energy, $\langle H_ {\rm em}\rangle/(\hbar \omega_c)  = \langle A^\dagger A\rangle  + 1/2$, for the same parameters. Note that for better visibility we have included for this plot the offset $\hbar \omega_c/2$ (indicated by the dashed line) into the definition of $H_{\rm em}$. For all plots $N=6$, $J = 0$ and $\gamma/\omega_c=0.04$ have been assumed. 
	}
	\label{fig:bbr}
\end{figure}

In Fig.~\ref{fig:bbr}(a) we plot $S_{\rm bb}(\omega)$ as a function of the coupling strength $g$ and for different temperatures. For small couplings, $g\ll \omega_c$, we see the expected splitting of the unperturbed cavity resonance into two polaritonic resonances at frequencies $\omega_\pm\approx \omega_c\pm G/2$. Although the lower polariton mode has a higher thermal occupation, the upper branch is slightly brighter. This observation can be partially explained by the scaling of the emission rate $\sim\omega_{\pm}^2$, but a more detailed analysis is presented below. At intermediate coupling strengths and temperatures the spectrum becomes rather complex. This is related to the large spread of the eigenenergies $E_n$ for these coupling values (see, for example, Fig. 1 in Ref.~\cite{Armata2017}) and the fact that the dipoles and photons are still strongly hybridized. At very large interactions, the spectrum collapses again to a single line centered around the bare cavity frequency. This collapse is a striking consequence of the approximate factorization of the eigenstates at very large interactions~\cite{Jaako2016} and provides a clear signature of the non-perturbative coupling regime, which can be detected in the emitted radiation field. While in Fig.~\ref{fig:bbr}(a) we plot the spectrum only for the case of non-interacting dipoles, the same overall trend is found independently of the dipole-dipole interaction strength $J$.  

Note that in previous studies of the absorption and emission spectra of the EDM~\cite{Todorov2014,Chervy2018} or the thermal radiation spectrum of the Rabi model ($N=1$)~\cite{Ridolfo2013c} only moderate values of $g$ have been considered, where this spectral collapse  does not yet occur. In the case of the Rabi model, it has also been shown that the statistics of the emitted photons can be sub-Poissonian for a certain range of couplings and temperatures~\cite{Ridolfo2013}. We don't find such a behavior for  larger numbers of two-level dipoles. 

\subsection{Radiated power and EM energy}

In Fig.~\ref{fig:bbr}(b) we also plot the total radiated power, $P_{\rm rad}$, and compare it with the equilibrium value of the EM field energy, $\langle H_{\rm em}\rangle$. Here, the total power is obtained from Eq.~\eqref{eq:Spectrum} by integrating over all frequencies,
 \begin{equation}\label{eq:Prad}
 P_{\rm rad}=  \frac{\hbar \kappa}{\omega_c}  \sum_{n>m} \frac{e^{-\beta E_n}}{Z} \omega_{nm}^2 |\langle E_n| A+A^\dag |E_m\rangle |^2.
 \end{equation}
For an empty cavity, this expression reduces to $P^0_{\rm rad}= \hbar \omega_c \kappa N_{\rm th}$, which we use to normalize the power values. 
Interestingly, for moderate temperatures we find a rather counterintuitive behavior: While the average energy that is stored in the mode increases for intermediate couplings, the emitted power decreases at the same time.  

\begin{figure}
\centering
	\includegraphics[width=\columnwidth]{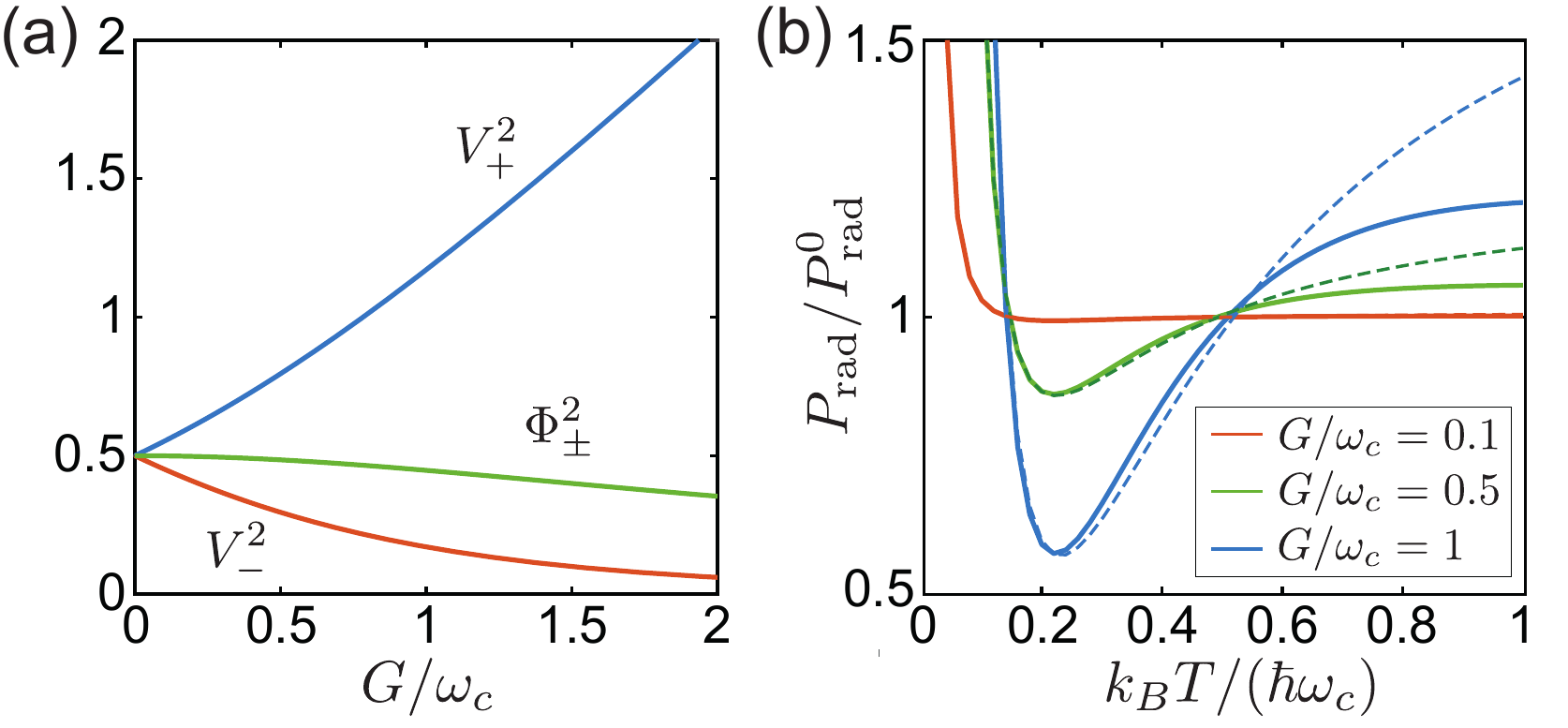}
	\caption{(a) The dimensionless matrix elements $V_\pm$ and $\Phi_\pm$, which determine the decomposition of the voltage and the magnetic flux operators in terms of the polariton operators $c_{\pm}$ [see Eqs.~\eqref{eq:VHP} and~\eqref{eq:PhiHP}] are plotted as a function of the collective coupling strength, $G$. (b) Plot of the ratio between the total power emitted from the coupled cavity QED system ($P_{\rm rad}$) and from the bare cavity ($P^0_{\rm rad}$) as a function of temperature. The solid lines are obtained from exact numerical calculations for $N=6$ and the dashed lines show the corresponding results predicted by Eq.~\eqref{eq:PowerHP} based on a Holstein-Primakoff approximation.}   
	\label{fig:Power}
\end{figure}

To explain this behavior we consider moderate values of $G\lesssim \omega_c$ and low temperatures. In this case we can use a Holstein-Primakoff approximation~\cite{HolsteinPrimakoff} and replace the spin operators $\sigma_-^i$ by bosonic annihilation operators $c_i$. The resulting linearized Hamiltonian can then be diagonlized and written as $H_{\rm cQED}\simeq H_{\rm HP}-\hbar \omega_c/2-N\hbar\omega_0$, where
\begin{equation}
 H_{\rm HP}=\sum_{\eta=\pm}  \hbar \omega_\pm \left(c^\dag_\pm c_\pm +\frac{1}{2}\right)+\sum_{k=1}^{N-1}  \hbar  \omega_0 \left(c_k^\dag c_k+\frac{1}{2}\right). 
\end{equation}
Here the $c_\pm$ are bosonic operators for the two bright polariton modes with frequencies $\omega_\pm$. The other bosonic operators $c_k$ represent dark polaritons, i.e.,  collective excitations of the dipoles, which are decoupled from the cavity due to symmetry. In terms of these polariton operators we can write
\begin{eqnarray}
A^\dag +A &=& V_+ \left(c_+^\dag +  c_+\right) + V_- \left(c_-^\dag  +  c_-\right)\!, \label{eq:VHP} \\
(A^\dag -A) &=& \Phi_+ \left(c_+^\dag -  c_+\right) +   \Phi_- \left(c_-^\dag  -  c_-\right)\!, \label{eq:PhiHP}
\end{eqnarray}
where the dimensionless matrix elements $V_\pm$ and $\Phi_\pm$ are plotted in Fig.~\ref{fig:Power}(a) as a function of the collective coupling strength $G$.

Within the Holstein-Primakoff approximation only the bright polariton modes contribute to the emission spectrum and we obtain
\begin{equation}\label{eq:PowerHP}
\frac{P_{\rm rad}}{P_{\rm rad}^0}  \simeq V^2_+ \left(\frac{\omega_+^2}{\omega_c^2}\right)   \frac{N_{\rm th}(\omega_+)}{N_{\rm th}(\omega_c)}+ V^2_-  \left(\frac{\omega_-^2}{\omega_c^2}\right) \frac{N_{\rm th}(\omega_-)}{N_{\rm th}(\omega_c)}.
\end{equation} 
We see that there are various competing effects. With increasing coupling $G$, the frequency $\omega_+$ and the matrix element $V^2_+$ for the upper polariton mode goes up, while at the same time the corresponding mode occupation, $N_{\rm th}(\omega_+)$, gets exponentially suppressed. The opposite is true for the lower polariton mode. As shown in Fig.~\ref{fig:Power}(b), this competition leads to a non-monotonic influence of the light-matter coupling on the radiated power. For temperatures $k_BT/(\hbar \omega_c)\approx 0.2-0.5$, as considered in Fig.~\ref{fig:bbr}(b), Eq.~\eqref{eq:PowerHP} indeed predicts the observed decrease in $P_{\rm rad}$ for increasing values of $G$. However, for higher and lower temperatures the dependence can also be reversed. In particular, for $k_BT/(\hbar \omega_c)\ll 1$ the occupation number of the bare cavity mode is exponentially suppressed. However, for $G\sim \omega_c$, also the lower polariton frequency is strongly reduced and $N_{\rm th}(\omega_-)\approx  k_BT/(\hbar \omega_-)$.  Under such conditions we observe a huge coupling-induced enhancement of the radiated power, $P_{\rm rad}/P_{\rm rad}^0\gg 1$.

By expressing also the EM energy, $H_{\rm em}=\hbar \omega_c A^\dag A$, in terms of the mode operators for the bright polariton modes we obtain 
\begin{equation}\label{eq:HEM_HP}
\begin{split}
\frac{\langle H_{\rm em}\rangle}{\hbar \omega_c}= & \frac{(V_+^2+\Phi_+^2)}{2} N_{\rm th}(\omega_+) + \frac{(V_-^2+\Phi_-^2)}{2} N_{\rm th}(\omega_-)\\
&+\left(\frac{V_+^2+\Phi_+^2+V_-^2+\Phi_-^2}{4}-\frac{1}{2}\right).
\end{split}
\end{equation}
We see that the prefactors for the thermal contributions in the first line of this equation have a much weaker dependence on $G$. Further, we find that for $G\gtrsim \omega_c$ and for temperatures $k_BT/(\hbar \omega_c)\lesssim 0.5$, the main contribution to the EM energy comes from a positive vacuum term given in the second line of Eq.~\eqref{eq:HEM_HP}. 
This part does not contribute to the radiated power such that overall $P_{\rm rad}$ and $H_{\rm em}$ display a very different dependence on $G$.

\section{Conclusions}
\label{sec:conclusions}

In summary, we have analyzed the basic thermal properties of cavity QED systems in the USC regime. By using various analytic approximations and exact numerical results for a moderate number of two-level dipoles, we have derived the coupling-induced corrections to the free energy and some of its derivative quantities. In the collective USC regime our analytic results confirm the basic intuition that the coupling to a single cavity mode cannot significantly change the properties of a larger ensemble of dipoles. While a large collective coupling strength $G$ has a substantial influence on the emission spectrum and also on the total radiated power, the corrections to material properties are small and scale only with the single-dipole coupling strength as $\sim g^2/\omega_c$. This means that major modifications of ground-state chemical reactions or cavity-induced shifts of ferroelectric phase transitions cannot be simply explained by a strong collective coupling to a single quantized mode. To identify  the detailed origin of such effects further experimental and theoretical investigations are still required, in particular also on the influence of the metallic boundaries on electrostatic interactions~\cite{DeBernardis2018,Vukic2012,Schuler2020,Bratkovsky2008} and other modifications of the background EM environment~\cite{Peters2019}. 

The behaviour of cavity QED systems changes completely once the single-dipole coupling parameter, $g/\omega_c$, becomes of order unity. For this regime we have shown in terms of several explicit examples that the coupling to a single cavity mode can strongly modify material properties such as the specific heat, the susceptibility or the ferroelectric phase transition temperature. While this regime is currently not accessible with atomic, molecular or excitonic cavity QED systems, such coupling conditions can be reached in superconducting quantum circuits, where also the effect of temperature is more relevant than in the optical regime. Apart from realizations with non-interacting~\cite{Jaako2016} or collective ferroelectric~\cite{Bamba2016} qubits, such setups would also allow to engineer $H_{\rm cQED}$ with arbitrary local qubit-qubit interactions by adding to those circuits additional capacitances or SQUID loops~\cite{Jaako2019}.
Therefore, such artificial light-matter systems constitute an ideal testbed for studying strongly-coupled quantum systems with unconventional thermodynamical properties. Potentially, this can also lead to more accurate  descriptions of fundamental thermodynamical processes or the optimization of quantum thermal machines, for which cavity QED and collective spin models have already been considered as simple toy systems in the past~\cite{Fusco2016,Ma2017,Cottet2017,Naghiloo2018,Masuyama2018,Alcalde2019}.

\acknowledgements
This work was supported by the Austrian Academy of Sciences (\"OAW) through a DOC Fellowship (D.D.) and a Discovery Grant (P.P., P.R.) and by the Austrian Science Fund (FWF) through the DK CoQuS (Grant No.~W 1210) and Grant No. P31701 (ULMAC).

\appendix

\section{Numerics} \label{app:Numerics}
To perform an exact numerical evaluation of the partition function and of other quantities derived from it we make use of the fact that $[H_{\rm cQED}, \vec S^2] = 0$. This means that  the Hamiltonian is block-diagonal in the eigenbasis of the collective spin operator, $\vec S=(S_x,S_y,S_z)$, and we can evaluate the partition functions $Z_s$ for each subspace separately, taking the multiplicity of each angular-momentum sector into account [see Eqs.~\eqref{eq:Z} and \eqref{eq:multiplicity}]. This reduces the dimension of the spin part from $2^N$ to $N +1$ states. For the actual calculation of $Z_s$ we first change into the polaron frame, as explained in Sec.~\ref{sec:npregime}. The Hamiltonian $\tilde H_{\rm cQED}$ is then projected onto the collective spin eigenstates $|s,m\rangle$, where $m=-s,\dots, s$, and the Hilbertspace of the photon mode is truncated at a maximal photon number $N_{\rm ph}$. 
For this truncated Hamiltonian we compute the full set of eigenvalues $\epsilon_s^i$ to obtain $Z_s=\sum_i e^{-\beta \epsilon_s^i}$. The cutoff number $N_{\rm ph}$ is increased until the results converge. Since the polaron transformation absorbes any mean displacements of the photon field, the actual cutoff numbers remain moderately large, even in the ferroelectric phase. For the parameter regimes considered in this paper we find that the heuristic choice $N_{\rm ph} = {\rm max}\{40, 20\times  \lceil k_B T_{\rm max}/(\hbar \omega_c) \rceil \}$ already produces very accurate results. If not mentioned otherwise, this value of $N_{\rm ph}$ is used for the plots presented in this paper.

\section{Mean-field theory for the LMG model}\label{app:LMG_MF}

For $J<0$ the LMG model constitutes a simple model for ferroelectricity with all-to-all interactions. It is thus expected that for $N\gg1$ the phase transition point is accurately predicted by the mean-field Hamiltonian
\begin{equation}
H_{\rm LMG}^{\rm MF} = \hbar \omega_0 S_z + \hbar \frac{2J}{N} \Sigma_x S_x - \hbar \frac{J}{N}\Sigma_x^2, 
\end{equation}   
where $\Sigma_x=\langle S_x\rangle$. Under this approximation, the resulting free energy is given by 
\begin{equation} \label{eq:F_LMG_MF}
F_{\rm LMG}^{\rm MF}(\Sigma_x)= -k_BT N \ln \left[ 2\cosh\left( \frac{ \hbar \Omega}{2k_BT}\right)\right] -  \hbar \frac{J}{N}\Sigma_x^2,
\end{equation}
where $\Omega=\sqrt{\omega_0^2 + 4J^2\Sigma_x^2/N^2}$. In the paraelectric phase the free energy has only a single minimum at $\Sigma_x=0$, while the ferroelectric phase is characterized by the appearance of two degenerate minima at a finite value of $\Sigma_x$. The transition between the two phases is given by Eq.~\eqref{eq:TcJc}.

Equation (\ref{eq:F_LMG_MF}) allows us to derive some useful relations for the spin expectation values of the LMG model. From the condition $\partial F_{\rm LMG}^{\rm MF}/\partial \Sigma_x = 0$ we find
\begin{equation}\label{eq:Sx_cond}
\frac{\partial F_{\rm LMG}^{\rm MF}}{\partial \Sigma_x} = -\frac{2\hbar J}{N} \left[  \frac{J }{ \Omega} \tanh \left(\frac{\hbar \Omega}{2 k_B T} \right) +1\right]\Sigma_x  = 0.
\end{equation}
For $J < J_c$ a solution $\Sigma_x \neq 0$ exists.
Subsequently, an expression for $\Sigma_z={\rm Tr} \{S_z \rho_{\rm LMG}^{\rm MF}\} $ can be derived via
\begin{equation}\label{eq:Sz}
\Sigma_z = \frac{\partial F_{\rm LMG}^{\rm MF}}{\partial (\hbar \omega_0)}  =-\frac{ N\omega_0}{2\Omega} \tanh \left(\frac{\hbar \Omega}{2 k_B T} \right).
\end{equation}
Eq.~\eqref{eq:Sx_cond} and Eq.~\eqref{eq:Sz}  are transcendental and therefore, in general, no explicit expressions for $\Sigma_x$ and $\Sigma_z$ can be found.

\section{Perturbation theory}\label{app:Perturbation}
Consider a generic system with Hamiltonian $H=H_0+H_g$ and free energy $F=F_0+F_g$, such that $F_0$ is the free energy of the unperturbed Hamiltonian $H_0$. In this case we obtain  the following general expression for the coupling-induced part of the free energy,
\begin{equation}
e^{-\beta F_g}=  \langle \mathcal{T} e^{-\int_0^\beta d\tau \, H_g(\tau)} \rangle_0 = e^{-\beta \sum_n F_g^{(n)}}.
\end{equation}
Here, $\langle \cdot \rangle_0$ denotes the average with respect to the thermal state of $H_0$,  $H_g(\tau)=e^{\tau H_0} H_g e^{-\tau H_0}$ and $\mathcal{T}$ is the time-ordering operator along the imaginary time axis. In systems where $H_g\sim g$ is only a small perturbation to $H_0$ we can use a cumulant expansion to approximate the expectation value in terms of a finite number of $F_g^{(n)}\sim g^n$.

\subsection{Weak coupling limit}  \label{App:Perturbation_g}
We first apply this perturbation theory to the coupling Hamiltonian $H_g$ as defined in Eq.~\eqref{eq:Hg}. Since in this case the thermal average of the linear term vanishes,  $\langle a+a^\dag\rangle_0=0$, the lowest-order correction to the free energy  is given by
\begin{equation}\label{eq:F2Integrals}
\begin{split}
F_g^{(2)}&=\frac{\hbar g^2}{\omega_c} \langle S_x^2\rangle_0 \\
&- \frac{\hbar^2 g^2}{\beta} \int_0^\beta d\tau_1\int_0^{\tau_1}d\tau_2 \, C(\tau_1,\tau_2) \langle S_x(\tau_1)S_x(\tau_2)\rangle_0,
\end{split}
\end{equation} 
where
\begin{equation}
\begin{split}
C(\tau_1,\tau_2) =\,&\langle[ a(\tau_1)+a^\dag(\tau_1)][a(\tau_2)+a^\dag(\tau_2)]\rangle_0\\
= \,& (N_{\rm th}+1) e^{-\hbar \omega_c(\tau_1-\tau_2)} +N_{\rm th} e^{\hbar \omega_c(\tau_1-\tau_2)}.
\end{split}
\end{equation}
For interacting dipoles the spin expectation value must be evaluated numerically. In terms of the energies $E_n$ and eigenstates $|E_n\rangle$ of $H_{\rm dip}$ we obtain 
\begin{equation}\label{eq:SxSx} 
\begin{split}
\langle S_x(\tau_1)S_x(\tau_2)\rangle_0= \frac{1}{Z^0_{\rm dip}} \sum_{n,m} &e^{-\beta E_n+(\tau_1-\tau_2)(E_n-E_m)} \\
&\times  |\langle E_n|S_x|E_m\rangle|^2.
\end{split}
\end{equation}
For non-interacting dipoles this calculation can be carried out analytically using 
\begin{equation}
\begin{split}
\langle S_x(\tau_1)S_x(\tau_2)\rangle_0=&\frac{1}{4} e^{(\tau_1-\tau_2)\hbar \omega_0} \langle S_+S_-\rangle_0\\
+ &\frac{1}{4}e^{-(\tau_1-\tau_2)\hbar \omega_0} \langle S_-S_+\rangle_0
\end{split}
\end{equation} 
and
$\langle S_\pm S_\mp\rangle_0= N \left[ 1  \mp \tanh\left(\hbar \omega_0/2k_BT\right)\right]/2$.
Altogether and writing $F_g^{(2)}=N\hbar g^2/(4\omega_c)f_g$ we obtain
\begin{equation}
\begin{split}
N f_g=N & - (N_{\rm th}+1) \langle S_+S_-\rangle_0 I(\omega_0-\omega_c)\\
&- (N_{\rm th}+1) \langle S_-S_+\rangle_0 I(-\omega_0-\omega_c)\\
&- N_{\rm th} \langle S_+S_-\rangle_0 I(\omega_0+\omega_c)\\
&-N_{\rm th} \langle S_-S_+\rangle_0 I(\omega_c-\omega_0),
\end{split}
\end{equation}
where
\begin{equation}
I(\Delta)= \frac{\hbar \omega_c}{\beta} \int_0^\beta d\tau_1\int_0^{\tau_1}d\tau_2 \, e^{\hbar \Delta(\tau_1-\tau_2)}.
\end{equation}
After some further simplifications the expression for $f_g$ reduces to the result given in Eq.~\eqref{eq:fgfull}.

\subsection{Bound on the free energy correction}
From the general expression for the correlation function given in Eq.~\eqref{eq:SxSx} one can show that  $\langle S_x(\tau_1)S_x(\tau_2)\rangle_0\leq \langle S_x^2\rangle_0$ for $\tau_1\geq\tau_2$. This inequality can be used together with $(N_{\rm th}+1)I(-\omega_c)+N_{\rm th} I(\omega_c)=1$ in the second line of Eq.~\eqref{eq:F2Integrals} in order to derive the upper and lower bounds $0\leq f_g\leq 4\langle S_x^2\rangle_0/N$. To improve the upper bound we can repeat the whole perturbation calculation in a displaced frame, $\tilde H_{\rm cQED} =\mathcal{D}^\dag(\alpha) H_{\rm cQED} \mathcal{D}(\alpha)=H_0+\tilde H_g(\alpha)$, where $\mathcal{D}(\alpha)=e^{\alpha a^\dag-\alpha^* a}$ is the displacement operator. Specifically, by choosing $\alpha=-g \langle S_x\rangle_0/\omega_c$ we obtain 
\begin{equation}
\tilde H_g\left(\alpha= -\frac{g\langle S_x\rangle_0}{\omega_c}\right) =  g(a+a^\dag) \Delta S_x + \frac{g^2}{\omega_c}(\Delta S_x)^2,  
\end{equation}
where $\Delta S_x= S_x-\langle S_x\rangle_0$. This means that we can repeat the analysis from above with $S_x$ being replaced by $ \Delta S_x$. This leads to the stricter bound for $f_g$ given in Eq.~\eqref{eq:fg2Bound}.

\subsection{Low-frequency limit} \label{App:Perturbation_w0}
We can use the same perturbation scheme to evaluate the lowest-order corrections to the free energy when $\omega_0 \rightarrow 0$. To do so we first change to the polaron frame as described in Sec.~\ref{sec:npregime} and decompose the total Hamiltonian as $\tilde H_{\rm cQED}=H_0 + H_{\omega_0}$. Here
\begin{equation}
\begin{split}
	H_0 & = \hbar \omega_c a^{\dag} a + \frac{\hbar}{4}\sum_{i,j} J_{ij} \sigma_x^i\sigma_x^j,
\\
	H_{\omega_0} & = \hbar \omega_0\left( \cos(\hat{\theta} ) S_z - \sin(\hat{\theta} ) S_y  \right),
\end{split}
\end{equation}
where  $\hat \theta= i (g/\omega_c)(a^\dag -a)$. According to this partitioning, the bare Hamiltonian $H_0$ is diagonal in the $\sigma_x$  basis and $\langle H_{\omega_0}\rangle_0=0$. Thus, up to second order in $\omega_0$ the free energy is given by $F \simeq F_0 + F_{\omega_0}^{(2)}$, where
\begin{equation}
\begin{split}
 F_{\omega_0}^{(2)} =  - \frac{\hbar^2 \omega_0^2}{\beta} &\int_0^{\beta} d\tau_1 \int_0^{\tau_1} d \tau_2 \, \\
& \times [ C_c (\tau_1,\tau_2) \langle{ S_z(\tau_1 ) S_z(\tau_2) \rangle}_0  \\
&\,\,\,\,\,+C_s (\tau_1,\tau_2) \langle{ S_y(\tau_1 ) S_y(\tau_2) \rangle}_0 ] ,
\end{split}
\end{equation}
where $C_c (\tau_1,\tau_2) $ and $C_s (\tau_1,\tau_2)$ are the correlation functions of the operators $\cos(\hat{\theta})$ and $\sin(\hat{\theta})$, respectively.
The correlation function for the photons can be calculated exactly using the properties of the displacement operator. Specifically, given a pair of complex number $z_1$ and $z_2$ the following formula holds
\begin{equation}\label{eq:ExpectDisplacement}
\langle  e^{z_1 a^{\dag} - z_2 a } \rangle_0  = e^{-z_1z_2(1 + 2 N_{\rm th} )/2}.
\end{equation}
The two-point correlation functions can then be expressed in terms of an infinite series,
\begin{equation}
C_c (\tau_1,\tau_2) =  e^{-\frac{g^2}{\omega_c^2}(1+2N_{\rm th})} \sum_{r,q=0}^{\infty} K_{rq} e^{(q-r)\omega_c (\tau_1-\tau_2) }
\end{equation}
and 
\begin{equation}
C_s (\tau_1,\tau_2) = e^{-\frac{g^2}{\omega_c^2}(1+2N_{\rm th})} \sum_{r,q=0}^{\infty} Q_{rq}  e^{(q-r)\omega_c (\tau_1-\tau_2) },
\end{equation}
with coefficients 
\begin{equation}
\begin{split}
K_{rq} & = \frac{\left[ 1 + (-1)^{r+q} \right]}{2} \left(\frac{g}{\omega_c}\right)^{2(r+q)} \frac{(1+N_{{\rm th}})^r N_{{\rm th}}^q}{r! ~ q!},
\\
Q_{rq} & = \frac{\left[ 1 - (-1)^{r+q} \right]}{2} \left(\frac{g}{\omega_c}\right)^{2(r+q)} \frac{(1+N_{{\rm th}})^r N_{{\rm th}}^q}{r! ~ q!}.
\end{split}
\end{equation}
Altogether we obtain 
\begin{equation}
\begin{split}
&F^{(2)}_{\omega_0} = - \frac{\hbar^2\tilde{\omega}_0^2}{\beta} \sum_{r,q=0}^{\infty} \int_0^{\beta} d\tau_1 \int_0^{\tau_1} d\tau_2  \, e^{(q-r)\omega_c (\tau_1-\tau_2) }  \\
&\times  \left[ K_{rq} \braket{S_z(\tau_1) S_z(\tau_2)}_0+ Q_{rq}\braket{S_y(\tau_1) S_y(\tau_2)}_0\right],
\end{split}
\end{equation}
where $\tilde{\omega}_0 = \omega_0 \exp[-g^2/(2\omega_c^2)(1+2N_{\rm th})]$. 

In Fig.~\ref{fig:Chiz}(a) this result is used to evaluate $\chi_z$ for non-interacting dipoles and, as expected, we find that it is in perfect agreement with the full numerical simulations for arbitrary $g$ and arbitrary temperatures. Note that in the series above one can interpret each term as a process in which $q$-photons are absorbed and $r$-photons are emitted. In this way it is clear that in the ultrastrong coupling regime there are processes with multiphoton scattering, emission and absorption. These processes become more and more important as the coupling strength increases.

\section{Variational LMG Hamiltonian}\label{app:Variational}
In the variational ansatz described in Sec. \ref{sec:npregime}, the trial Hamiltonian $H^*$ given in Eq.~\eqref{eq:Hstar} is simply the decoupled cavity QED system with $\omega_0$ being replaced by a renormalized frequency $\tilde \omega_0$. Therefore, we obtain $F^*=F_c^0+F_{\rm dip}^0(\tilde \omega_0)$ and 
\begin{equation}
\langle H-H^*\rangle_{\rho^*} =  \hbar \omega_0 \langle \cos ( \hat \theta) S_z \rangle_0 - \hbar\tilde \omega_0 \langle S_z\rangle_0,
\end{equation}
where we have already assumed $\langle S_y\rangle_0=0$. The thermal expectation value of the cosine operator can be evaluated with the help of Eq.~\eqref{eq:ExpectDisplacement}.

Altogether, the variational free energy $F_V$ can be written as
\begin{equation} \label{eq:F_V1}
F_V= F_c^0+ F_{\rm dip}^0(\tilde \omega_0) +\hbar\left( \omega_0e^{-\frac{g^2}{\omega_c^2} (N_{\rm th} +1/2)  }  - \tilde \omega_0 \right)\langle S_z\rangle_0.
\end{equation}
Taking the derivative of $F_V$ with respect to $\tilde \omega_0$ yields the extremal condition
\begin{align}
\frac{\partial F_V}{\partial \tilde \omega_0} = &\frac{\partial F_{\rm dip}^0(\tilde \omega_0) }{\partial \tilde \omega_0}  - \hbar \langle S_z\rangle_0 \\
&+ \hbar\left( \omega_0e^{-\frac{g^2}{\omega_c^2} (N_{\rm th} +1/2)  }  - \tilde \omega_0 \right) \frac{\partial \langle S_z\rangle_0}{\partial \tilde \omega_0} \overset{!}{=}0. \nonumber
\end{align}
Using the general relation $\partial F_{\rm dip}^0(\tilde \omega_0) / \partial ( \hbar \tilde \omega_0 ) =  \langle S_z\rangle_0$ we obtain
\begin{equation}
\tilde \omega_0 = \omega_0 e^{-\frac{g^2}{2\omega_c^2}(1+2N_{\rm th} ) } .
\end{equation}
Note that this result is independent of the dipole-dipole couplings $J_{ij}$.

\section{Emission spectrum}\label{app:Emission}
For $C_t\ll C$, the capacitive coupling between the cavity and the transmission line shown in Fig.~\ref{Fig1:Setup} can be modeled by a Hamiltonian of the form $H_{\rm c-t}=C_t V V_t$. Here $C_t$ is the coupling capacitance and $ V_t=\sum_k V_k (b^\dag_k +b_k)$ is the voltage operator  of the transmission line, which we expressed in terms of a set of free EM modes with bosonic operators $b_k$ and frequency $\omega_k$. We write $V=V_0 X$ and define $\lambda_k= C_tV_0V_k/\hbar$ such that the total Hamiltonian of the cavity QED system and the transmission line reads
\begin{equation}
H= H_{\rm cQED} + \sum_k \hbar \omega_k b_k^\dag b_k +  \hbar \sum_k \lambda_k (b^\dag_k + b_k) X . 
\end{equation}
For a conventional transmission line we have $\lambda_k\sim \sqrt{\omega_k}$.

We can formally integrate the equations of motion for the mode operators $b_k$,
\begin{equation}
b_k(t) = b_k(0)e^{-i\omega_k t} -i   \lambda_k \int_{0}^t dt'  \,e^{-i\omega_k(t-t')} X(t').
\end{equation}
Therefore, by assuming that initially the transmission line is in the vacuum state, we find that 
\begin{equation}
\braket{b_k^{\dag}b_k}(t) = \lambda_k^2 \int_{0}^t\int_{0}^t d t' d t'' \, e^{i\omega_k(t''-t')}\braket{X(t')X(t'')}.
\end{equation}
To proceed we write $X(t)=X_-(t)+X_+(t)$, such that $X_+(t)= \sum_{n\geq m}  X_{nm} e^{i\omega_{nm} t}$ contains only contributions that oscillate with a positive frequency, $\omega_{nm}=(E_n-E_m)/\hbar\geq 0$, and $X_-(t)=X^\dag_+(t)$~\cite{Ridolfo2013}. Then, for long enough times $t\gg \gamma^{-1}$, where $\gamma$ is the characteristic thermalization rate of the  cavity QED system, we obtain 
\begin{equation}
\langle b_k^{\dag}b_k\rangle(t) \simeq  \lambda_k^2  t \times 2 {\rm Re} \int_{0}^\infty d\tau  \, \braket{X_+(\tau)X_-(0)}_0 e^{-i\omega_k \tau }.
\end{equation}

The total power radiated into the transmission line is 
\begin{equation}
 P_{\rm rad}= \sum_k  \hbar \omega_k \partial_t \langle b_k^{\dag}b_k\rangle(t)
\end{equation}
and by writing $ P_{\rm rad}= \int_0^\infty d\omega\, S_{\rm bb}(\omega)$ we obtain the general expression for the black-body spectrum 
\begin{equation}
S_{\rm bb}(\omega)= \hbar \omega J(\omega) C_{X}(\omega). 
\end{equation}
Here $J(\omega)=\sum_k \lambda_k^2 \delta(\omega-\omega_k)$ is the spectral density of the transmission line and 
\begin{equation}
C_{X}(\omega) =2 {\rm Re} \int_{0}^\infty d\tau  \, \braket{X_+(\tau)X_-(0)}_0 e^{-i\omega \tau }.
\end{equation}
By assuming an Ohmic spectral density we can write $J(\omega)=\kappa \omega/(2\pi \omega_c)$, where $\kappa$ is the rate at which the bare cavity decays into the transmission line. For a completely isolated system the correlation function of the system operator $X$ is given by
\begin{equation}
C_{X}(\omega) = 2\pi \sum_{n>m} \frac{e^{-\beta E_n}}{Z} |\langle n| X |m\rangle |^2 \delta(\omega -\omega_{nm} ),
\end{equation}
and we obtain the total emitted power given in Eq.~\eqref{eq:Prad}. 

For the evaluation of the black-body spectrum $S_{\rm bb}(\omega)$ one must keep in mind that the cavity QED system is in constant interaction with the surrounding thermal bath, which induces a finite broadening of all the spectral lines. A detailed investigation of such line-broadening effects is beyond the scope of the current analysis. Instead, we simply replace all  resonances by a Lorentzian profile with a phenomenological width $\gamma$, while still approximating $\omega J(\omega) \simeq \omega_{nm} J(  \omega_{nm} )$ for each transition. As a result, we obtain the spectrum given in Eq.~\eqref{eq:Spectrum}.

%

\end{document}